\documentclass[twocolumn]{aastex631} 

\usepackage[T1]{fontenc}
\usepackage{color,epsfig,graphics,graphicx,rotating,comment,amssymb,stackengine}
\usepackage{natbib,lipsum,atbegshi}
\usepackage{natbib,lipsum,atbegshi,lineno}
\usepackage{sidecap}
\usepackage{ulem}
\usepackage{natbib}
\usepackage{hyperref}
\usepackage{todonotes}
\usepackage{rotating} % provides \begin{sidewaysfigure*} environment for Fig. 2 (RGA)

\definecolor{brown}{rgb}{0.6,0.4,0.2} 
\definecolor{purple}{rgb}{0.5,0,0.5} 

\shorttitle{JWST NIR observations of Cas A} 
\shortauthors{Rho et al.}

\def\one{{\,\sc i}}
\def\two{{\,\sc ii}}

\newcommand{\kms}{km\,s$^{-1}$}

\newcommand{\spitzer}{\textit{Spitzer}}

\newcommand{\ariif}{\ion{[Ar}{2}]}

\newcommand{\mic}{$\mu$m}

\newcommand{\herschel}{\textit{Herschel}}
\newcommand{\new}{\textcolor{black}}

\shorttitle{Carbon Monoxide in Cas A with JWST}

\begin{document}

\title{Shockingly Bright Warm Carbon Monoxide Molecular Features in the Supernova Remnant Cassiopeia A Revealed by JWST}  

\author{J. Rho}\affil{SETI Institute, 189 Bernardo Ave., Ste. 200, Mountain View, CA 94043, USA; jrho@seti.org} \affil{Department of Physics and Astronomy, Seoul National University, Gwanak-ro 1, Gwanak-gu, Seoul, 08826, South Korea}
\author{S.-H. Park}\affil{Department of Physics and Astronomy, Seoul National University, Gwanak-ro 1, Gwanak-gu, Seoul, 08826, South Korea}
\author{R. Arendt}\affil{Center for Space Sciences and Technology, University of Maryland, Baltimore County, Baltimore, MD 21250, USA}
\affil{Code 665, NASA/GSFC, 8800 Greenbelt Road, Greenbelt, MD 20771, USA}
\author{M. Matsuura}\affil{Cardiff Hub for Astrophysics Research and Technology (CHART), School of Physics and Astronomy, Cardiff University, The Parade, Cardiff CF24 3AA, UK}
\author{D. Milisavljevic}\affil{Department of Physics and Astronomy, Purdue University, 525 Northwestern Avenue, West Lafayette, IN 47907, USA}
\author{T. Temim}\affil{Department of Astrophysical Sciences, Princeton 
University, Princeton, NJ 08544}
\author{I. De Looze}\affil{Sterrenkundig Observatorium, Ghent University, Krijgslaan 281-S9, B-9000 Ghent, Belgium}
\author{W. P. Blair}\affil{William H. Miller III Department of Physics and Astronomy,
Johns Hopkins University, 3400 N. Charles Street, Baltimore, MD, 21218}
\author{A. Rest}\affil{Space Telescope Science Institute, 3700 San Martin Drive, Baltimore, MD 21218, USA}
\affil{William H. Miller III Department of Physics and Astronomy,
Johns Hopkins University, 3400 N. Charles Street, Baltimore, MD, 21218}
\author{O. Fox}\affil{Space Telescope Science Institute, 3700 San Martin Drive, Baltimore, MD 21218, USA}
\author{A. P. Ravi}\affil{Department of Physics and Astronomy, University of California, 1 Shields Avenue, Davis, CA 95616-5270, USA}
\author{B.-C. Koo}\affil{Department of Physics and Astronomy, Seoul National University, Gwanak-ro 1, Gwanak-gu, Seoul, 08826, South Korea} 
%\affil{Center for Research and Exploration in Space Science and Technology, NASA/GSFC, Greenbelt, MD 20771, USNASA/GSFc, UBD}
\author{M. Barlow}\affil{Department of Physics and Astronomy, University College London, Gower Street, London WC1E 6BT, United Kingdom}
\author{A. Burrows}\affil{Department of Astrophysical Sciences, Princeton 
University, Princeton, NJ 08544} % ; Orchid: 0000-0002-3099-5024}
\author{R. Chevalier}\affil{Department of Astronomy, University of Virginia, Charlottesville, VA 22904-4325, USA}
\author{G. Clayton }\affil{Department of Physics \& Astronomy, Louisiana State University, Baton Rouge, LA 70803, USA}
\author{R. Fesen}\affil{6127 Wilder Lab, Department of Physics and Astronomy, Dartmouth College, Hanover, NH 03755, USA}
\author{C. Fransson}\affil{Department of Astronomy, Stockholm University, The Oskar Klein Centre, AlbaNova, SE-106 91 Stockholm, Sweden}
\author{C. Fryer}\affil{Center for Nonlinear Studies, Los Alamos National Laboratory, Los Alamos, NM 87545 USA}
\author{H. L. Gomez}\affil{Cardiff Hub for Astrophysical Research and Technology (CHART), School of Physics \& Astronomy, Cardiff University, The Parade, Cardiff, CF24 3AA, UK}
\author{H.-T. Janka}\affil{Max Planck Institute for Astrophysics, Karl-Schwarzschild-Str. 1, 85748 Garching, Germany}
\author{F. Kirchschlarger}\affil{Sterrenkundig Observatorium, Ghent University, Krijgslaan 281-S9, B-9000 Ghent, Belgium}
\author{J. M. Laming}\affil{Space Science Division, Code 7684, Naval Research Laboratory, Washington DC 20375, USA}
\author{S. Orlando}\affil{INAF - Osservatorio Astronomico di Palermo, Piazza del Parlamento 1, 90134 Palermo, Italy} 
\author{D. Patnaude}\affil{Center for Astrophysics | Harvard \& Smithsonian, 60 Garden St, Cambridge, MA 02138, USA}
\author{G. Pavlov}\affil{Pennsylvania State University, Department of Astronomy \& Astrophysics, 525 Davey Lab., University Park, PA 16802, USA}
\author{P. Plucinsky}\affil{Center for Astrophysics | Harvard \& Smithsonian, 60 Garden St, Cambridge, MA 02138, USA}
\author{B. Posselt}\affil{Oxford Astrophysics, University of Oxford, Denys Wilkinson Building, Keble Road, Oxford, OX1 3RH, UK} 
\author{F. Priestley}\affil{Cardiff Hub for Astrophysical Research and Technology (CHART), School of Physics \& Astronomy, Cardiff University, The Parade, Cardiff, CF24 3AA, UK}
\author{J. Raymond}\affil{Center for Astrophysics | Harvard \& Smithsonian, 60 Garden St, Cambridge, MA 02138, USA}
\author{N. Sartorio
}\affil{Sterrenkundig Observatorium, Ghent University, Krijgslaan 281-S9, B-9000 Ghent, Belgium}
\author{F. Schmidt}\affil{Department of Physics and Astronomy, University College London, Gower Street, London WC1E 6BT, United Kingdom} 
\author{P. Slane}\affil{Center for Astrophysics | Harvard \& Smithsonian, 60 Garden St, Cambridge, MA 02138, USA}
\author{N. Smith}\affil{Steward Observatory, University of Arizona, 933 N. Cherry Avenue, Tucson, AZ 85721, USA}
\author{N. Sravan}\affil{Department of Physics, Drexel University, Philadelphia, PA 19104, USA}
\author{J. Vink}\affil{Anton Pannekoek Institute for Astronomy \& GRAPPA, University of Amsterdam, Science Park 904, 1098 XH Amsterdam, The Netherlands}
\author{K. Weil}\affil{Department of Physics and Astronomy, 6127 Wilder Laboratory, Dartmouth College, Hanover, NH 03755, USA}
\author{J. Wheeler}\affil{University of Texas at Austin, 1 University Station C1400, Austin, TX 78712-0259, USA}
\author{S. C. Yoon}\affil{Department of Physics and Astronomy, Seoul National University, Gwanak-ro 1, Gwanak-gu, Seoul, 08826, South Korea}

\begin{abstract} 

We present JWST NIRCam (F356W and F444W filters) and MIRI (F770W) images and NIRSpec-IFU spectroscopy of the young Galactic supernova remnant Cassiopeia A (Cas A) to probe the physical conditions for molecular CO formation and destruction in supernova ejecta. We obtained the data as part of a JWST survey of Cas A. The NIRCam  and MIRI  images map the spatial distributions of synchrotron radiation, Ar-rich ejecta, and  CO on both large and small scales, revealing remarkably complex structures. The CO emission is stronger at the outer layers than the Ar ejecta, which indicates the reformation of CO molecules behind the reverse shock. NIRSpec-IFU spectra (3 – 5.5 $\mu$m) were obtained toward two representative knots in the NE and S fields that show very different nucleosynthesis characteristics.  Both regions are dominated by the bright fundamental rovibrational band of CO in the two R and P branches, with strong [Ar\,{\small VI}] and relatively weaker, variable strength ejecta lines of [Si\,{\small IX}], [Ca\,{\small IV}], [Ca\,{\small V}] and [Mg\,{\small IV}]. The NIRSpec-IFU data resolve individual ejecta knots and filaments spatially and in velocity space. The fundamental CO band in the JWST spectra reveals unique shapes of CO, showing a few tens of sinusoidal patterns of rovibrational lines with pseudo-continuum underneath, which is attributed to the high-velocity widths of CO lines. Our results with LTE modeling of CO emission indicate a temperature of $\sim$1080 K and provide unique insight into the correlations between dust, molecules, and highly ionized ejecta in supernovae, and have strong ramifications for modeling dust formation that is led by CO cooling in the early Universe.

\end{abstract} 

%ApJ doesn't need keyword (they use their own system to select keywords on line).

\section{Introduction}

The large amounts of dust seen in some high-$z$ galaxies imply that dust formed in the early Universe. However, intermediate-mass stars, thought to produce most interstellar dust when on the asymptotic giant branch (AGB) in present-day galaxies \citep{laporte17}, would not have evolved to the dust-producing stage in high-$z$ galaxies. In contrast, core-collapse supernovae (ccSNe) from high mass stars occur just several million years after their progenitors are born, and have also been suggested to be molecular factories in the early Universe \citep{cherchneff08}.  Molecules, e.g., carbon monoxide (CO), are the signature of the onset of dust formation since they are one of the ejecta's most powerful coolants \citep{cherchneff08} at temperatures where dust can form.

However, whether SNe are a significant (if not dominant) source of dust in the early universe has been and continues to be debated \citep[e.g.][]{nozawa06, cherchneff09}.  The dust masses observed in ccSNe within a few years after their explosions are less than 0.01\,M$_\odot$, a value that is far too small to explain the amount of dust observed in the early Universe \citep{kotak09,gall11, tinyanont19}. In contrast, \herschel\ and \spitzer\ observations of a few young supernova remnants (SNRs), including SN~1987A (20 \,yrs) \citep{matsuura11} and Cas A ($\sim$350\,yrs), have dust masses of 0.1--1 $M_\odot$ \citep[][and references therein]{rho18, millard21,matsuura15,chawner19, delooze17}, which is in agreement with dust formation models \citep{nozawa03,todini01, sluder18} that suggest that SNe could be major dust factories at high-$z$ galaxies \citep{dwek11}.

The cause of discrepancies in the measured dust masses from early to later phases and the timescale of dust formation are under debate. Recently, \cite{niculescu-Duvaz22, shahbandeh23} suggest that a dust mass grows with time, and most dust forms at times $>$ 3 yr after the explosion, while \cite{dwek19} suggests that the dust forms early in optically thick clumps and only a fraction of its IR emission is detected due to high IR opacity.

Dust may undergo complete or partial destruction following its initial formation, which depends on shock velocity and grain size \citep{slavin20}, potentially altering the expected dust grain size distribution in the process.  When the forward shock of a SN accumulates sufficient pressure at the SN shell, a second shock (called a reverse shock) develops interior \citep{chevalier77, borkowski90}. The dust (or ejecta) destruction by the passage through the reverse shock of ccSNe depends on grain composition, grain size distribution, and the shock properties \citep{priestley21,nozawa07,kirchschlager19,kirchschlager23}. Because the dust formed in ccSNe includes sufficiently large grains (0.1 - 0.5 \mic),  a significant fraction of the grains can survive \citep[10 - 20\% for silicate dust and 30 - 50\% for carbon dust;][]{slavin20}.

Cas A is one of the youngest ($\sim$350 yr) and closest ejecta-dominated SNRs, and has been observed across all wavelengths, including optical \citep{fesen06}, X-ray \citep{hwang04}, radio \citep{deLaney14}, and infrared \citep{Isensee10, Isensee12, smithJD07midIR}. The progenitor mass of Cas A is still uncertain; it has been suggested it was a Wolf-Rayet star with a mass of 15--25 $M_\odot$ \citep{fesen01, young06}. Spectra of a light echo from the SN that produced Cas A showed a Type IIb SN from the collapse of the helium core of a red supergiant that had lost most of its hydrogen envelope \citep{krause08}. \citet{koo20} alternatively suggest a blue supergiant precursor with a thin hydrogen envelope or a yellow supergiant (the progenitor mass is probably $<$15 $M_\odot$, depending on pre-supernova mass loss).

Cas A is the best case study in the local Universe to understand dust formation in SN ejecta and shock-processing of freshly formed SN dust \citep{delooze17}. {\it Spitzer} infrared spectral mapping of Cas A confirmed that molecules and dust are present in the ejecta \citep{rho08}. Cas A also shows strong polarization fractions ($\sim$ 20\%) in the far-infrared, implying that the dust grains are large \citep{rho23casapol}. CO was detected for the first time from the Palomar and \spitzer\ NIR images \citep{rho09, rho12}. CO fundamental band features were detected in low-resolution AKARI spectra, showing that astrochemical processes and molecule formation continues to the stage of young SNRs \citep{rho09, rho12}.

Although CO molecules themselves are not on the immediate chemical path to dust formation \citep{sarangi15}, it is known that the detection of CO is an indication of molecular cooling and chemistry in the ejecta, leading to condensation of ejecta dust in later (from a few tens to a few hundred days after the explosion) epochs \citep{sarangi18, rho18, rho21}. Although CO can form in the ejecta in early phases  \citep[$\sim$100\,days after SN explosion; ][]{sarangi13}, the detection of later epoch CO has been associated with reformation in the post-shock region \citep{biscaro14}, highlighting the complex and competing processes of molecular formation and destruction.

CO molecular lines have been observed in the late-time NIR spectroscopy of a number of supernovae, e.g., two Type IIP SN1987A~\citep{spyromilio88} and SN 2017eaw~\citep{rho18sn,tinyanont19} and a Type Ic SN2020oi \citep{rho21}. By comparing the production of molecules, ejecta, and dust at these early times to the molecular gas in Cas A, we can constrain both production and destruction of dust and molecules in supernova explosions.

In this paper, we present new observations of Cas A using JWST, focusing on NIRCam and MIRI imaging of the entire SNR and NIRSpec IFU observations of two selected filaments. JWST reveals that the CO-emitting regions mostly coincide spatially with the ejecta-dominated areas; however, the ratio between the CO and Ar ejecta varies across the SNR and fine-scale differences in position and morphology exist between the CO emission and the ejecta.  The NIRSpec spectra are dominated by complex CO bands and show a unique shape of the CO fundamental band.

\section{Observations}

We present JWST infrared observations of Cas A using NIRCam, MIRI, and NIRSpec. These observations were obtained as part of a Cycle 1 survey program on Cas~A (Prog.~ID of 1947), described in \cite{milisavljevic24}. The NIRCam images were taken on 2022 December 5-6, using the F356W  (3.140--3.980\,\mic) and F444W (3.880--4.986\,\mic) filters.  The field of view (FOV) of NIRCam is 2.2$'$ $\times$ 4.4$'$ with the pixel scale of 0.065$''$. In contrast, \spitzer\ images at these wavelengths have a spatial resolution of 3$''$ \citep{fazio04}. We required a 2$\times$3  mosaic to cover the full extent of Cas A. The NIRCam data required 1.9 hr on source exposure time for F345W and F444W images.

\begin{figure*}[p]
%\begin{largefigure}
%\begin{center}
\includegraphics[scale=0.5,angle=0,width=16.8truecm]{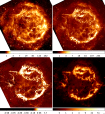}
\caption{JWST F356W (3.56 \mic; top left), F444W (4.44 \mic; top right) images and CO fundamental image (produced from F444W image after subtracting the synchrotron emission using F356W image; bottom left), and equivalent CO image using {\it Spitzer} data (bottom right). The JWST images show far more complexity. The CO emission is from the CO fundamental bands and overlaps spatially with the knotty ejecta structures. Faint extended, diffuse F444W emission seen in projection toward the center, hinted at with the {\it Spitzer} data, is more noticeable. 
The image is centered on RA\ $23^{\rm h} 23^{\rm m} 26.65^{\rm s}$ and Dec\ $+58^\circ$49$^{\prime}14.98^{\prime \prime}$ (J2000) with a FOV of 6.4$'$ $\times$ 6.4$'$. The units in the color bar are in MJy\,sr$^{-1}$.}
\label{jwstcompspitzer}
%\end{center}
%\end{largefigure}
\end{figure*}

\begin{sidewaysfigure*}
\begin{center}
\vspace{3.5in}
\includegraphics[width=9in]{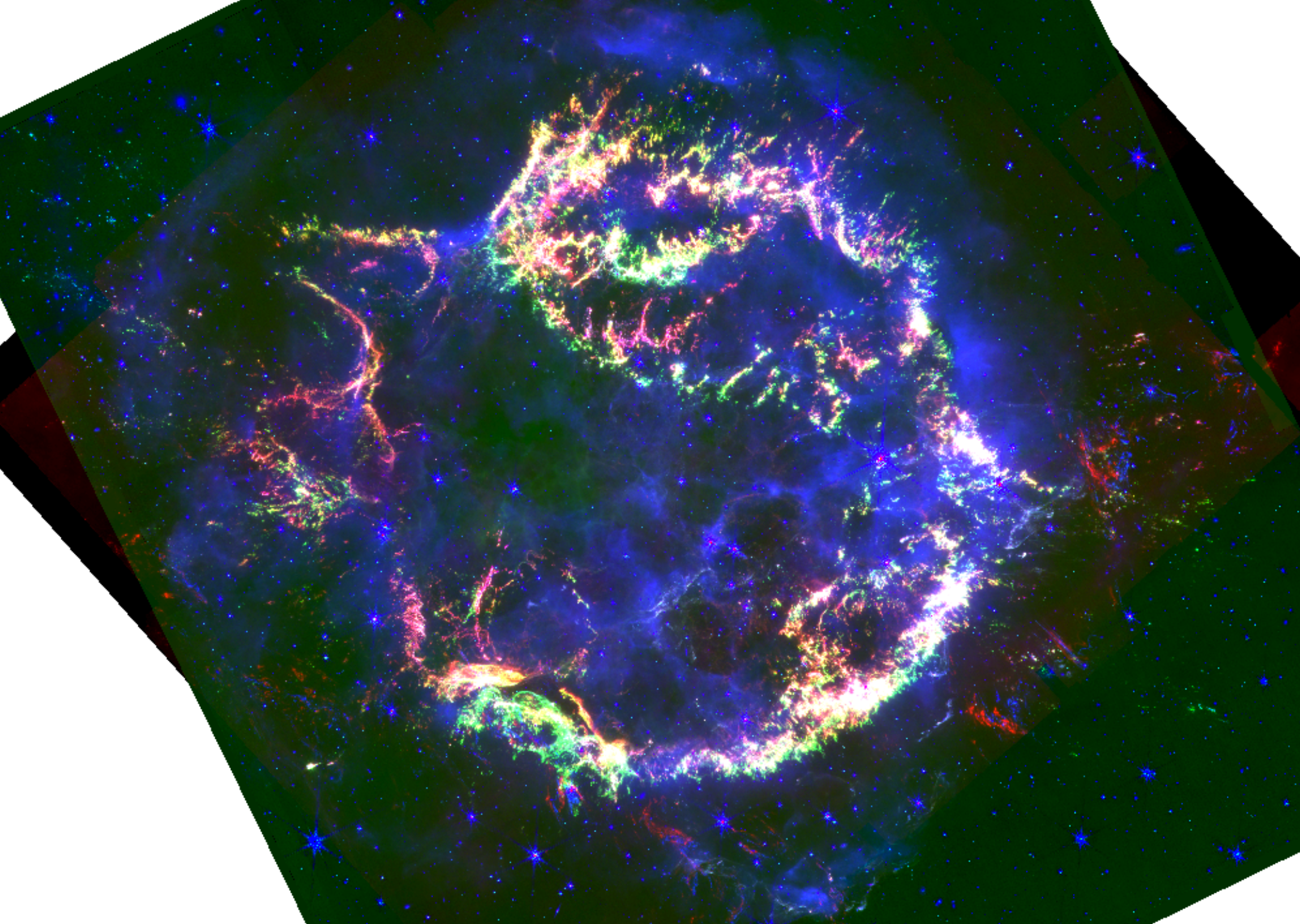}
%{Figures/jwstcasabF356gCOrF777nobarcut.pdf}
\caption{JWST three-color mosaicked images of synchrotron emission (F356W in blue), CO (synchrotron-subtracted F444W in green) and Ar ejecta (F770W in red). North is up and East is to the left. The detailed structures of the three images are noticeably different from each other. The synchrotron emission (blue) shows smooth structures, and is dominant outside the main shell. 
} 
\label{jwstcolor}
\end{center}
\end{sidewaysfigure*}

MIRI imaging using the F770W filter were taken multiple times between 2022 August 4 and October 26.  The MIRI imaging FOV is 1.23$'$ $\times$ 1.88$'$ with the pixel scale of 0.11$''$, requiring a   5$\times$3 map to cover the SNR. The JWST mosaics were astrometrically aligned using the JWST Alignment Tool (JHAT) \citep{rest23}. The resulting aligned mosaics are shown in color in Figures \ref{jwstcompspitzer} and \ref{jwstcolor}.

\begin{figure*}
\begin{center}
\includegraphics[scale=0.6,angle=0,width=16.5truecm]{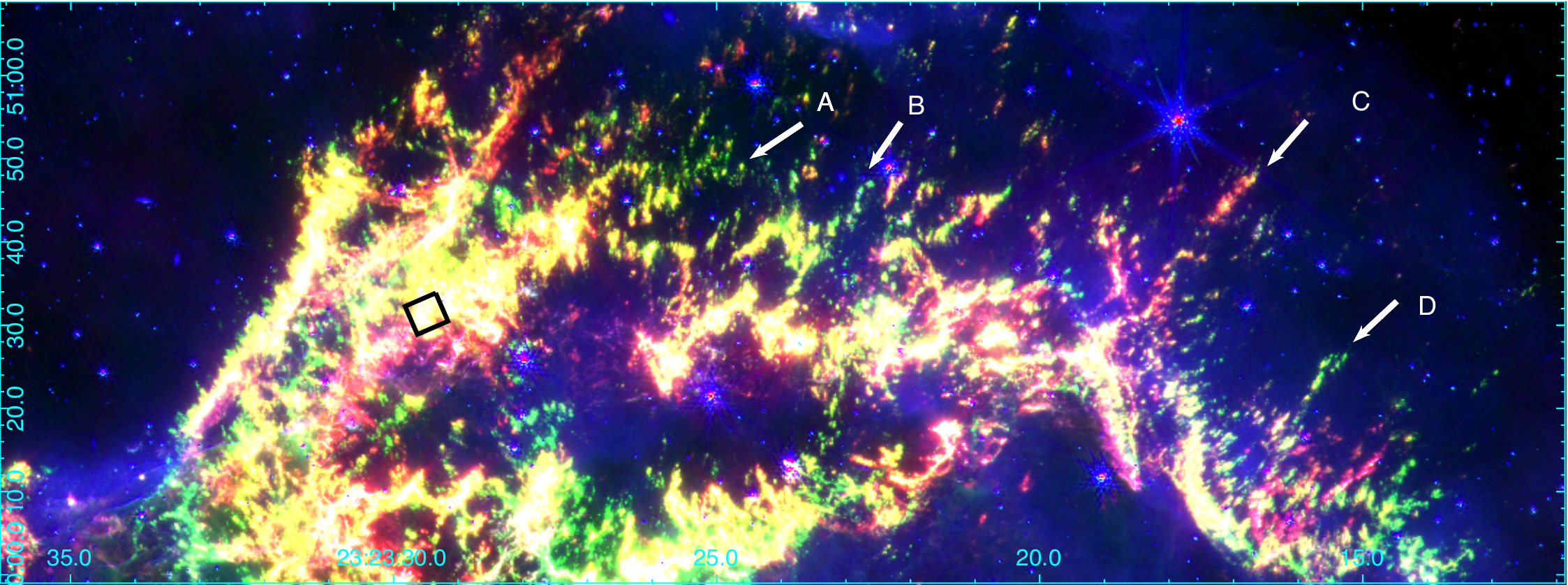}
%{Figures/jwstIFU08FOVcyanlabelsnew.pdf}
\includegraphics[scale=0.6,angle=0,width=16.5truecm]{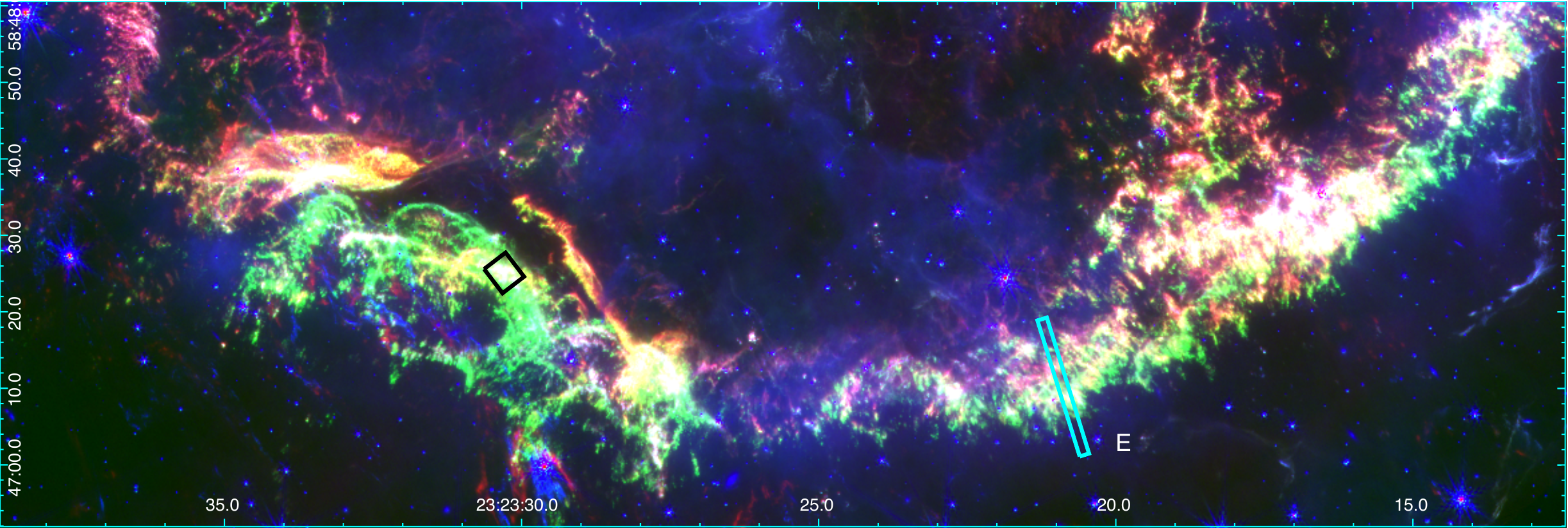}
%{Figures/jwstIFU10SFOVwhite.pdf}
\caption{Zoomed images  of Figure \ref{jwstcolor} on the northern \textit{(top}) and southern \textit{(bottom)} shells. The NIRSpec-IFU FOV are marked as black squares  on the JWST three-color images. The arrows point to the filaments showing CO excess emission (in green, marked as A, B, and D) and Ar ejecta excess (in red, marked as C). The slit (box E, in cyan) is cut through the southern shell where a radial profile is obtained in Figure~\ref{sbprofiles}.}
\label{jwstcolornorth}
\end{center}
\end{figure*}

Each NIRSpec-IFU position covers a region 3.7$''$x$4''$ in size with 0.1$''$ pixels. Four IFU positions were observed for the program, as described in \cite{milisavljevic24}. Here we focus on two positions that were observed, one on 2022 November 5 and the other on 2022 December 10.  The position in the north (R.A. = 350.873$^{\circ}$,  Decl. = 58.842$^{\circ}$) targeted an ejecta knot and the position in the south (R.A. = 350.875$^{\circ}$,  Decl. = 58.790$^{\circ}$) targeted a particularly bright region of CO emission.  These positions (P1 and P3 in \cite{milisavljevic24}) are shown in Figure \ref{jwstcolornorth}. The other two regions are reported by \cite{deLooze24, milisavljevic24}. We extracted the JWST NIRSpec-IFU spectra from each pixel, and the total $\sim$1368 spectra are from 36$\times$38  detector pixels. However, note that the IFU had 53$\times$55 image size after wcs rotation. The effective integration and total exposure times are 145 and 583 s for the N field and 218 and 875 s for the S field, respectively. More details of the JWST observations of Cas A are described in \cite{milisavljevic24}.

%Figure1

%Figure2

% Figure3

\section{Results}
%Figure4
\begin{figure}[!ht]
\begin{center}
\includegraphics[scale=0.6,angle=0,width=8.5truecm]{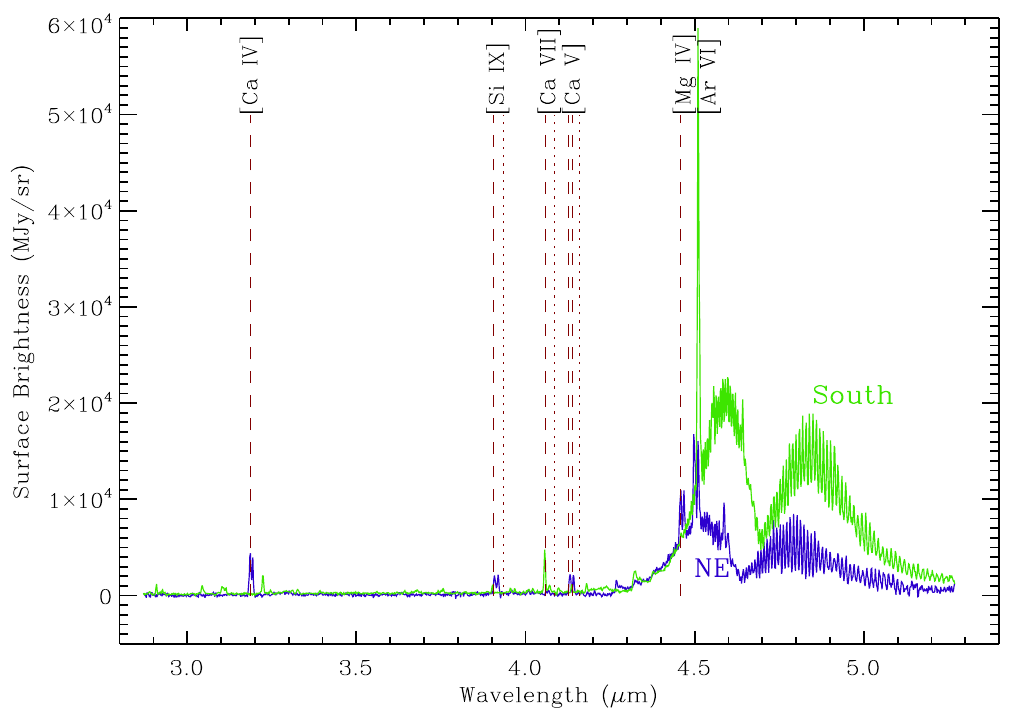}
%{Figures/jwstcasaCOspecNEaSbf5a.pdf}
\caption{JWST NIRSpec-IFU spectra of Cas A toward two representative knots (0.1$''$$\times$0.1$''$ region for each) in the northern IFU field (R.A.\,= 350.8731$^{\circ}$, Decl.\,= 58.8422$^{\circ}$ in blue) and
at the southern IFU field (R.A.\,= 350.8764$^{\circ}$, Decl.\,= 58.7905$^{\circ}$ in green). The spectra show a dominant CO fundamental band with a few high ionization ejecta lines indicated. The two CO bumps arise from the R (the shorter wavelength bump) and P (the longer wavelength bump) branches of CO.  The CO emission is stronger in the southern position, as expected from the imagery presented above. The ejecta lines are from [Ca~IV] at 3.2 \mic, [Si~IX] at 3.94 \mic, [Ca~V] at 4.16 \mic, [Mg~IV] at 4.49 \mic, [Ar~VI] at 4.53 \mic, and [K~III] at 4.6 \mic.}
\label{casanirspecCO}
\end{center}
\end{figure}

%Figure5
\begin{figure*}
\begin{center}
\includegraphics[scale=0.6,angle=0,width=8.5truecm]{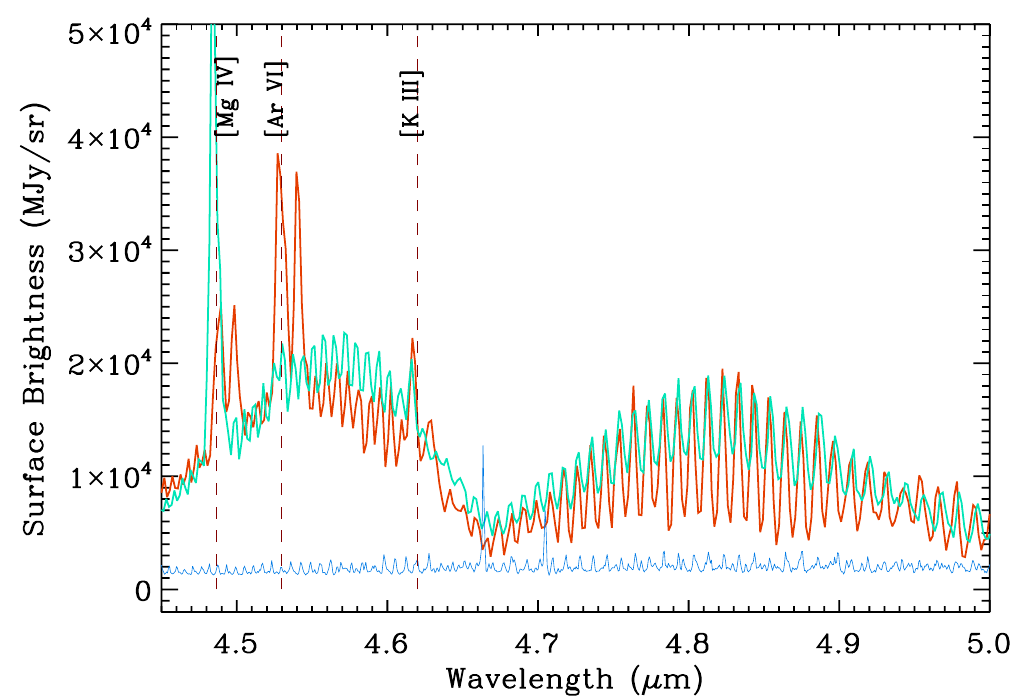}
%{Figures/decy23casajwstIFUspecorionbf6a.pdf}
\includegraphics[scale=0.6,angle=0,width=8.5truecm]{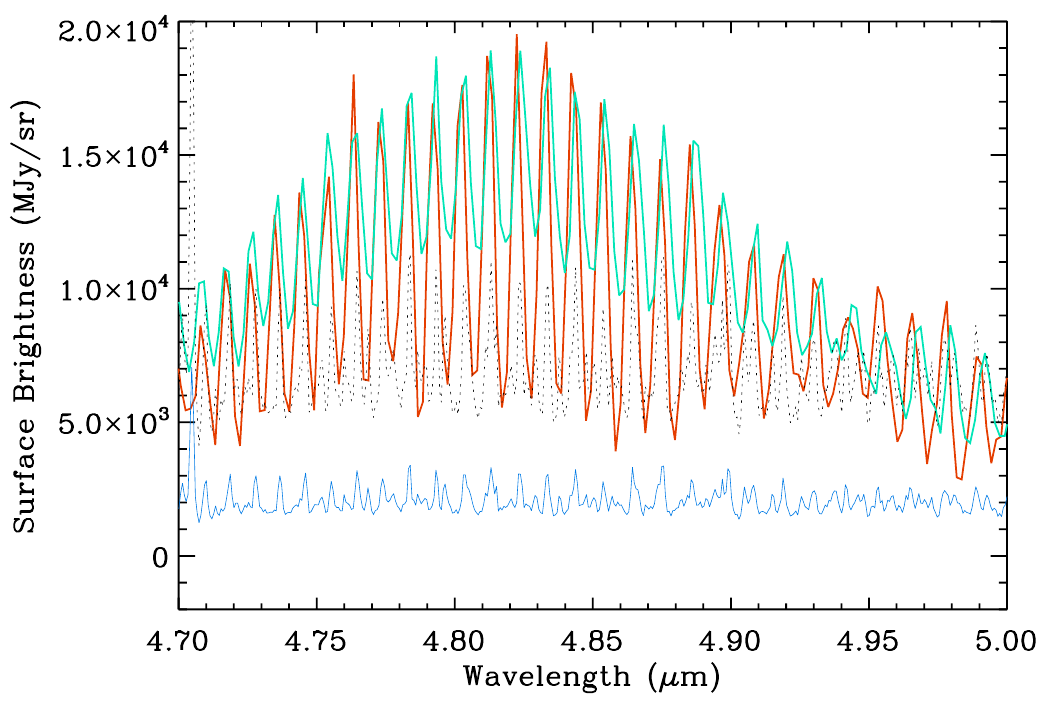}
%{Figures/casajwstspecorionzoomf6b.pdf}
\caption{\textit{(a: left)} Two representative NIRSpec-IFU spectra from the northern shell (in orange) and southern shell (green) show dominant features from CO fundamental bands with a few high ionization ejecta lines, in comparison with Orion (blue)  \cite{peeters23}.
The northern spectra show two velocity components of -2070 and -1270 \kms, while the southern spectra show one component of +1060 \kms. 
The $\sim$3400 \kms\ difference between the blue and red-shifted motion of Cas A has been removed using the rest wavelengths of the ejecta lines.
\textit{(b: right)} An enlarged section from the left figure shows a comparison of the lines with those of Orion (multiplied by 10, shown in dotted black.).
}
\label{casanirspecCOorion}
\end{center}
\end{figure*}

%Figure6
\begin{figure*}
\begin{center}
\includegraphics[scale=0.6,angle=0,width=16.truecm]{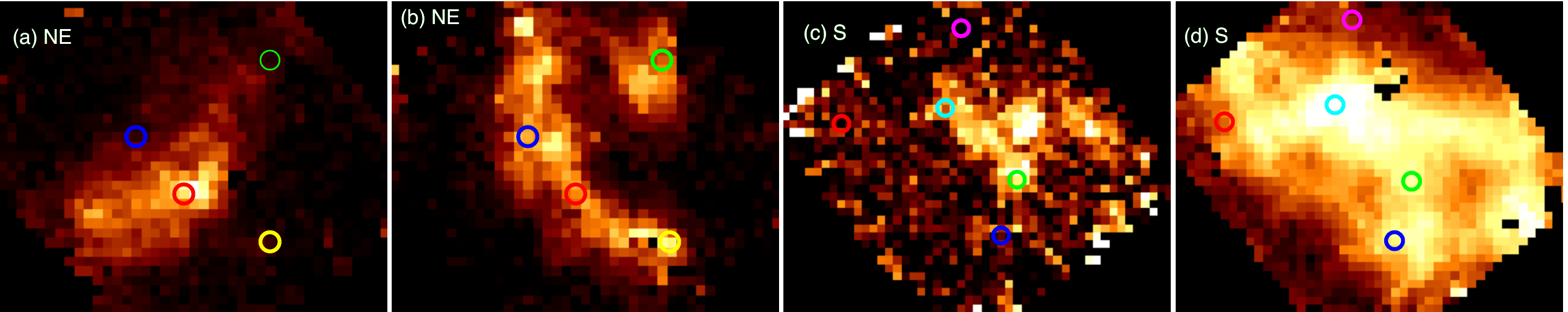}
%{Figures/jwstfourpanelregions4white.pdf}
\includegraphics[scale=0.6,angle=0,width=16.5truecm]
{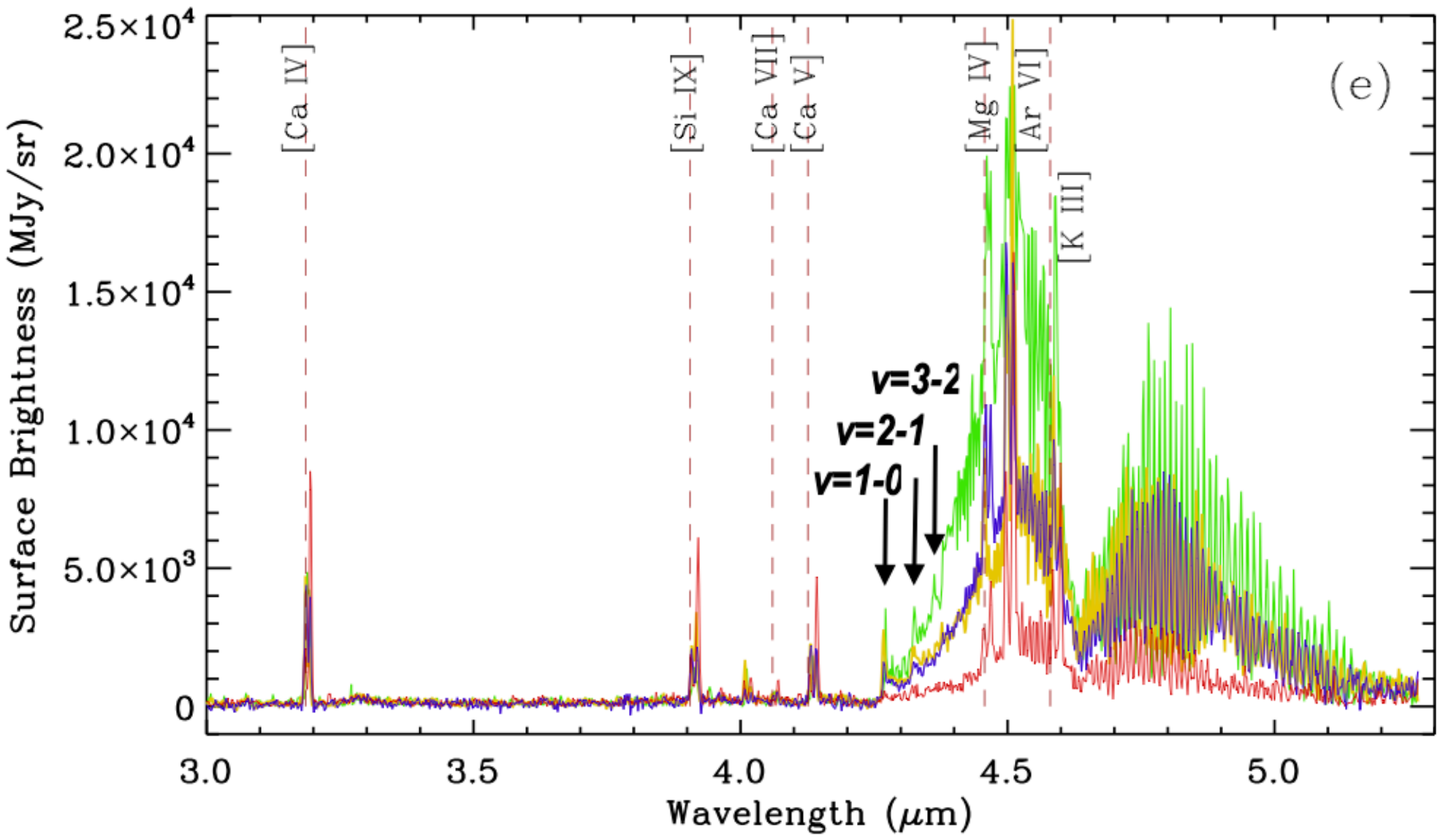}
%{Figures/IFUnorthmanyspec6labelstiff.pdf}
\includegraphics[scale=0.6,angle=0,width=16.5truecm]{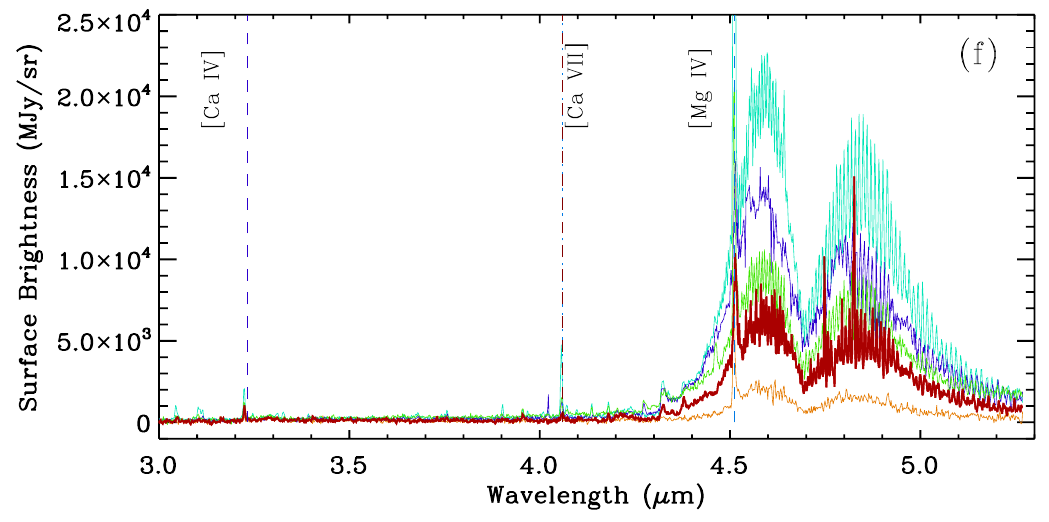}
%{Figures/IFUsouthmanyspec6.pdf}
\caption{\textit{(Top)} Regions for the extracted spectra shown below are marked on the images of NE IFU FOV for (a) 3.92 \mic\ ([Si IX]), (b) 4.70 \mic\ (CO), and of the S IFU FOV of (c) 3.92 \mic\ ([Si IX]) and (d) 5.04 \mic\ (CO). NIRSpec-IFU spectra of Cas A toward a few positions in the northern \textit{(middle: e)} and south \textit{(bottom: f)}. 
The positions and marked circles on (a) and (b) and extracted spectra (e) are (R.A., Decl.) = (350.8724$^{\circ}$, 58.8423$^{\circ}$) in green, (350.8732$^{\circ}$, 58.8421$^{\circ}$) in blue, (350.8729$^{\circ}$, 58.8419$^{\circ}$) in red, (350.8724$^{\circ}$, 58.8418$^{\circ}$) in yellow. The positions and marked circles on (c) and (d) and extracted spectra (f) are (R.A., Decl.) = (350.8763$^{\circ}$, 58.7907$^{\circ}$) in magenta (c, d) or orange (f), (350.8764$^{\circ}$, 58.7904$^{\circ}$) in cyan,  (350.8771$^{\circ}$, 58.7904$^{\circ}$) in red, (350.8759$^{\circ}$, 58.7902$^{\circ}$) in green, (350.8760$^{\circ}$, 58.7900$^{\circ}$) in blue.}
\label{casanirspecregions}
\end{center}
\end{figure*}

\subsection{Imaging}
In Figure \ref{jwstcompspitzer}, we present the mosaicked NIRCam images. The top two panels show the separate JWST NIRCam mosaics from F356W and F444W.  Much of the F356W emission appears relatively diffuse, similar to synchrotron radiation seen in the radio \citep{anderson91}, K-continuum \citep{rho03}, and \spitzer\ 3.6 \mic\ \citep{rho12} images. However, we note some knotty structures, likley ejecta, in the F356W image of the northern shell, which are visibly more compact and brighter than the synchrotron emitting regions. The NIRSpec spectra also show a continuum from synchrotron emission, but that at some locations the F356W band includes a [Ca~IV] 3.2 \mic\ line (see Figure \ref{casanirspecCO} and Section \ref{SIFU} for details). Thus, while there may be isolated, localized contributions from ejecta emission, the F356W image mainly shows synchrotron emission.

The F444W image show three main structures; (1)  knotty structures from ejecta, (2) the forward-shocked material outside the bright shell, which shows a much smoother distribution, and (3) weak diffuse structures in the interior (see Figure \ref{jwstcompspitzer}). The NIRSpec-IFU spectra show CO emission is dominant ($>$ 99\%) in compact knots and filaments, and that the F444W filter also includes ejecta lines of [Si~IX] at 3.94 \mic, [Ca~V] at 4.16 \mic, [Mg~IV] at 4.49 \mic, [Ar~VI] at 4.53 \mic, and [K~III] at 4.6 \mic\ (see Section \ref{SIFU} for details). The F444W filter also includes synchrotron emission since the forward-shocked material is known to be from synchrotron emission, as seen in radio images \citep{anderson91}.

To remove the synchrotron component to first order and obtain a CO image, we scaled the F356W image to the F444W image assuming a synchrotron spectral index $\alpha$ = $-$0.68 \citep{rho03} where log $S_\nu$ $\propto$ $\alpha$ log $\nu$ and subtracted it from the F444W image, resulting in the lower left panel in Figure \ref{jwstcompspitzer}.  This image is dominated by the inner ring of emission and should be due primarily to fundamental CO band emission, as previously suggested using \spitzer\ IRAC band 4 (at 4.5 \mic) imagery and AKARI spectra \citep{rho12}; this will be confirmed and discussed further using NIRSpec in the next section.  The fourth panel in Figure \ref{jwstcompspitzer} shows the \spitzer\ CO image for comparison \citep{rho12}.  Individual CO structures in the \spitzer\ image are resolved into numerous knots in the JWST images.

With this information in hand, we show the resulting aligned mosaics in color in Figure \ref{jwstcolor}. Synchrotron (F356W) is shown in blue, fundamental CO emission (F444W after synchrotron subtraction) is shown in green, and the Ar ejecta emission (from \ariif\ at 6.98\mic) are shown in red. The Ar ejecta image is from the MIRI F770W data which cover 6.58 -- 8.687 \mic. The \spitzer\ spectrum shows that the Ar lines are dominant within these wavelengths \citep{rho08, smith09}. The synchrotron emission arises mainly from Fermi acceleration in the forward shock region as it sweeps up and encounters ISM/CSM \citep{rho03}. The three-color mosaicked images contrast synchrotron emission, CO molecules, and Ar ejecta emission in great detail on all scales all the way down to individual ejecta knots. While the ejecta and CO emissions both arise primarily in the bright ring of emission caused by the reverse shock, their  spatial distributions are quite different from each other. Here we highlight two regions: the so-called ``Ne-moon'' regions \citep[Fig. 8 of][]{smith09, ennis06} where the Ne emission is brighter than Ar emission; these are the two regions that appear in green and white in Figure \ref{jwstcolor}.

We find that the spatial distribution of Ar (eastern region in red of Figure \ref{jwstcolor}) and CO regions (northern and southern regions in green) is similar to the spatial distribution of different dust types identified by \citet[][see Figure 2f]{rho08}; silicate (21 \mic\ dust) and carbon and Al$_2$O$_3$ dust regions.  These  are zones of  different nucleosynthesis and hence correspondingly very different dust formation and composition. The CO regions are in He/O/C and O/C layers \citep[zone 4A and 4B,][]{sarangi13, sarangi15} and the Ar ejecta are in Si/O layers (zone 1B).

\subsection{IFU Near-Infrared spectroscopy}
\label{SIFU}
\begin{table*} [!htbp]
%\begin{table*}
\caption{Observed spectral line width and brightness in the two representative spectra}
\label{Tlinefluxes}
\begin{center}
\begin{tabular}{llcllccl}
\hline \hline
Region& Line & $\lambda_0$ &Wavelength &  FWHM  & vel. Shift & vel. width &  Line Brightness  \\
&      & (\mic)    & (\mic)  &(\mic)&(\kms)      &  (\kms)    & (10$^{-3}$\,erg\,s$^{-1}$\,cm$^{-2}$\,sr$^{-1}$)\\
\hline
N & [Ca IV]& 3.2067& 3.1862$\pm$0.0005  & 0.0059$\pm$0.0013 & -1918$\pm$46 & 552$\pm$122  & 7.970$\pm$1.790\\
  & [Ca IV]& 3.2067& 3.1937$\pm$0.0005 & 0.0033$\pm$0.0010 & -1216$\pm$47 & 309$\pm$94 & 3.948$\pm$1.196 \\
  \hline
N &[Si IX] &3.9357  &  3.9086$\pm$0.0019  & 0.0069$\pm$0.0047 & -2065$\pm$145 & 526$\pm$358& 2.669$\pm$1.839 \\
& [Si IX] &3.9357  &  3.9190$\pm$0.0016  & 0.0052$\pm$0.0038 & -1151$\pm$121 & 396$\pm$290 & 2.123$\pm$1.563 \\
\hline
N &[Ca V] & 4.1585 & 4.1315$\pm$0.0016  &0.0062$\pm$0.0038 & -1947$\pm$115& 447$\pm$275  &2.475$\pm$1.531 \\
  &[Ca V] & 4.1585 & 4.1419$\pm$0.0013 &0.0043$\pm$0.0032   &-1197$\pm$94& 310$\pm$230& 1.709$\pm$1.261  \\
\hline
N &[Mg IV]&4.4866 & 4.4583$\pm$0.0019 &0.0061$\pm$0.0049& -1892$\pm$70 & 407$\pm$327 &5.230$\pm$4.109 \\
  &[Mg IV]&4.4866 &4.4683$\pm$0.0019 &0.0047$\pm$0.0046 &-1223$\pm$127 & 314$\pm$307 &3.587$\pm$3.515\\
  \hline
N &[Ar VI]& 4.5295&4.4990$\pm$0.0012& 0.0073$\pm$0.0028 &-2020$\pm$80 & 483$\pm$185 & 11.160$\pm$4.363 \\
  &[Ar VI]&4.5295& 4.5105$\pm$0.0010 & 0.0049$\pm$0.0024 &-1258$\pm$1192 & 324$\pm$150 & 7.048$\pm$3.505    \\
\hline
N &[K III] & 4.6180 & 4.5870$\pm$0.0046 & 0.0046$\pm$0.0044 & -2013$\pm$278 &298$\pm$285  & 3.365$\pm$4.970\\ 
&[K III] & 4.6180 &4.5985$\pm$0.0067 & 0.0070$\pm$0.0170 &-1266$\pm$435 & 454$\pm$304 & 2.594$\pm$6.279\\ 
\hline \hline
S&\new{UID}$^a$&... &   3.1055$\pm$0.0001&    0.0086$\pm$0.0003&... &830$\pm$10$^b$&2.020$\pm$0.083 \\
S&[Ca IV]& 3.2067&    3.2236$\pm$0.0000&  0.0050$\pm$0.0001& +1581$\pm$09 & 468$\pm$10 & 2.922$\pm$0.125\\
S&\new{UID}$^a$&... & 3.7547$\pm$0.0002&    0.0148$\pm$0.0006& ... &1182$\pm$48$^b$ &0.861$\pm$0.052\\
%  & \new{or [Mg\,I]}$^a$ & 3.7325 &3.7547$\pm$0.0002&    0.0148$\pm$0.0006& +1526$\pm$10 &1182$\pm$48$^c$ &0.861$\pm$0.052\\
S&  \new{UID}$^a$ &...  &  3.9024$\pm$0.0001& 0.0035$\pm$0.0002&...&263$\pm$15& 0.642$\pm$0.045 \\
S&\new{UID}$^a$&...&    4.0570$\pm$0.0000&    0.0050$\pm$0.0001& ... &365$\pm$8&4.082$\pm$0.163\\
S&\new{CO v=1-0}&...&    4.3248$\pm$0.0002&    0.0100$\pm$0.0004& ...&634$\pm$284$^c$&3.053$\pm$0.144\\   
S & [Mg IV] & 4.4866 &    4.5109$\pm$0.0001 &    0.0063$\pm$0.0002& +1619$\pm$07 &421$\pm$13 & 39.080$\pm$1.955\\
S &[K III] & 4.6180 &   4.6421$\pm$0.0003&    0.0035$\pm$    0.0006& +1598$\pm$18 & 231$\pm$39 &3.021$\pm$0.707 \\
\hline
\end{tabular}
\end{center}
\renewcommand{\baselinestretch}{0.8}
{\hskip 0.3truecm}\footnotesize{$^a$ UID indicates an unidentified line. $^b$ The lines are broader, likely due to a contribution from another faint line(s). $^c$ The line is broader because of contributions from a few J transitions.}
\end{table*}

The positions of the NIRSpec-IFU apertures are shown in Figure \ref{jwstcolornorth} and representative extracted NIRSpec-IFU spectra are shown in Figures~\ref{casanirspecCO} and \ref{casanirspecCOorion}. These demonstrate the dominant nature of CO emission between 4--5 \mic\ as well as the presence of other emission lines. The strongest ejecta lines are from [Ca~IV] at 3.2 \mic, [Si~IX] at 3.94 \mic, [Ca~V] at 4.16 \mic, [Mg~IV] at 4.49 \mic, [Ar~VI] at 4.53 \mic, and [K~III] at 4.6 \mic. The fundamental CO features are very bright from 4.2 - 5.3 \mic\ and reveal unique shapes of CO, showing a few tens of sinusoidal patterns of rovibrational lines with pseudo-continuum underneath. There are many other faint lines that are yet to be identified, which is out of the scope of this paper. There are strong lines of [Ar~VI] 4.53 \mic\ on the top of the CO feature. We also note the wavelength range within the MIRI F770W image is dominated by  [Ar~II] at 6.985 \mic, which is 100 - 1000 times stronger than the continuum \citep{ennis06, rho08, smith09}.

The strongest detected lines and their properties are summarized in Table 1 with derived velocity information. In contrast with the previous \spitzer\ data where multiple velocity components were blended together, JWST resolves the knots in velocity space. The northern field shows pairs of blue-shifted emission lines at velocities of $-$2000 \kms\ and $-$1200 \kms. The southern field shows a single velocity component at +1600 \kms. The observed Doppler velocity structures are consistent with the 3-D plot (Doppler velocity vs projected radius) of ejecta from the \spitzer\ Infrared Spectrograph observations \citep[][]{Isensee12,deLaney10}. However, the JWST spectra reveal two reasonably well-separated Doppler-shifted groups of ejecta layers at -2000 and -1200  \kms\ in the northern filaments and a group of knots at +1600 \kms\ in the southern filament.  Interestingly, all observed elements of Mg, Si, Ar, K, and Ca ejecta show similar velocity components.

A combination of the velocity shift and width indicates different nucleosynthetic layers. The two velocity shift components at -1200 and -2000 \kms\ for the NE region are likely caused by the different nucleosynthetic layers \citep[see discussion by][]{Isensee12}. However, the velocity width may be only a factor of 2-3 times larger than the instrumental spectral resolution ($\sim$111 \kms), and yet its error is large (20 - 100 \%). There may be some separation between these layers by a few hundred \kms\ (corresponding to $\sim$5$''$ or $\sim$0.1 pc), and the width may indicate the physical size of the layer. For example, [Ca IV] layers are separated almost by $\sim$800 \kms\ ($\sim$0.3 pc) between -2000 and -1200 \kms\ layers with each of $\sim$0.1 pc size, while the separation of [K III]  layers is unclear due to the significant errors in the velocity widths.

We compare the NIRSpec spectra between the north and south regions in Figure \ref{casanirspecCOorion}. When we shift the spectra to align the features, we see a velocity shift of roughly 3400 \kms\ between the north and south. The north is blue-shifted compared with the south, and consistent with previous 3D reconstruction of the remnant \citep{deLaney10,milisavljevic13}. We also compared the forest of CO features in Cas A with those of the Orion nebula. Cas A's CO features are bright with prominent individual covibrational features and show a broad underlying component, which we attribute to kinematic and instrumental broadening (see Section \ref{SLTE} for details). JWST spectra of the protoplanetary disk HH 48 also show CO features similar to Orion, but in absorption \citep{sturm23}.

We compare the NIRSpec spectra from a few regions. In Figure \ref{casanirspecregions} we show spectra extracted from various sub-regions within the IFU fields. The locations of the spectral extraction regions are shown in the top panels Figure \ref{casanirspecregions}a-\ref{casanirspecregions}d and the color-coded spectra are shown in the panels below. The overall intensity and the contrast of the CO lines show variation, and variable strength ejecta lines are seen. The northern position spectra show more prominent ejecta lines than the southern region spectra.\\

\section{Discussion} 
\subsection{Ejecta and CO maps}

\begin{figure}
%\begin{center}
\includegraphics[scale=0.6,angle=0,width=8truecm]{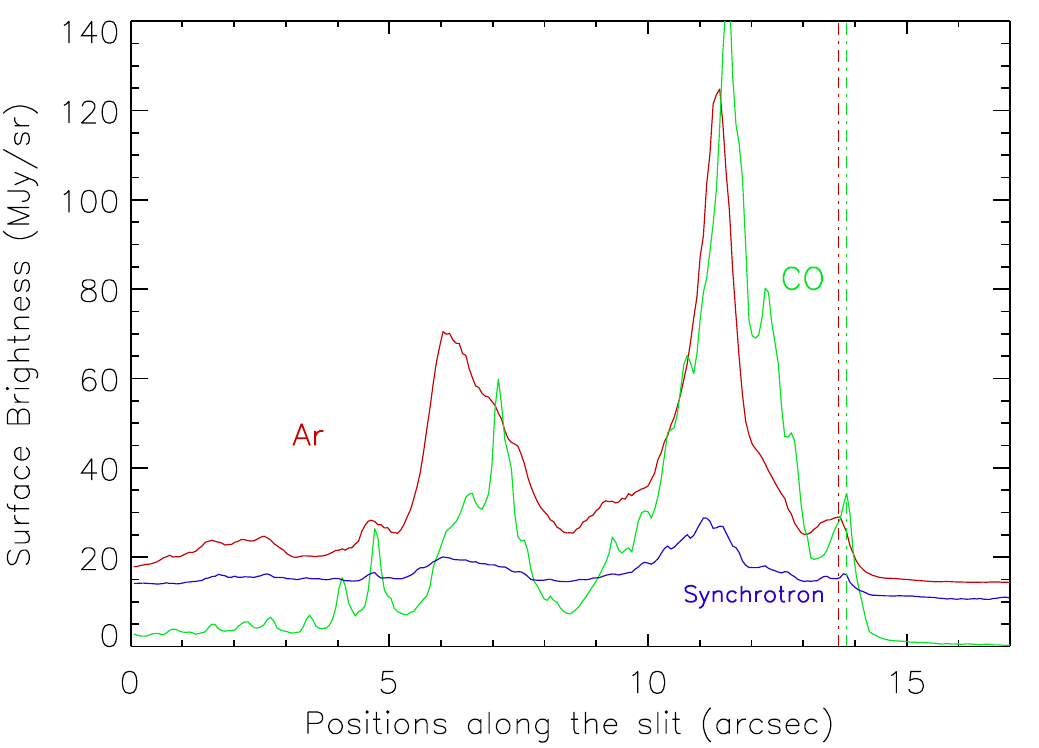}
%{Figures/jwstsbprofilesarcsec3.pdf}
\caption{Profiles of surface brightness in Ar ejecta, CO, and synchrotron emission along a slit that cuts through the southern shell in an approximately radial direction (the slit is marked as box E in Figure~\ref{jwstcolornorth}b). The position extends from (R.A., Decl.) = (350.838$^{\circ}$, 58.789$^{\circ}$) to (350.836$^{\circ}$, 58.784$^{\circ}$). The CO peaks further outward than the Ar emission. The vertical lines are peaks of Ar (in red) and CO emission (in green), respectively.}
\label{sbprofiles}
\end{figure}

We have examined the distribution of Ar, CO, and synchrotron emission, shown in Figure~\ref{jwstcolor}. Figure \ref{jwstcolornorth} shows an enlargement toward the bright northern and southern limbs, where CO emission can often be seen exterior to the Ar ejecta emission. Some examples of regions where enhanced CO emission lies at larger radii from the explosion center are marked with arrows A, B, and D in Figure \ref{jwstcolornorth}, and box E in Figure \ref{jwstcolornorth}. We have examined the CO and Ar emission morphology and found almost one-to-one correspondence between their emission, but CO emission is often shifted to appear at larger radii than the Ar emission. We quantified this using a radial profile cut across the southern shell (marked as Region E in Figure \ref{jwstcolornorth}). The radial profiles of the cut across the southern shell in Figure \ref{sbprofiles} show that the peak in CO (dotted line in green) is located radially outward from that in Ar (dotted line in red). Synchrotron emission (in a blue curve) shows a weaker correlation with Ar ejecta and CO emission as its morphology differs from Ar and CO maps and knots.

Two scenarios are possible for the observation that the CO peak is offset outward from the Ar ejecta. The first scenario is that the CO molecules are re-forming behind the reverse shock, as suggested by \cite{biscaro14}, as the shock propagates inwards. 
The other possibility is simply that Ar ejecta has cooled faster than CO, causing a lack of infrared emission of the Ar ejecta in the outskirts. In the reverse shock, the initial shock velocity is low, and the gas density is high, so cooling may occur behind the shock wave \citep{chevalier77}. Since the ejecta are at fairly high temperatures ($>$ 10$^6$ K), they may cool faster than the CO molecules.

However, the reverse shock moves rapidly as the density of expanding gas decreases, so the shock wave soon heats the gas to a temperature at which its cooling time is long. Only a few percent of the ejected mass may be cooled after the reverse shock has passed, as we observe ejecta knots in the far outer, high-velocity part of the shell \citep{fesen01}. Most ejecta in the reverse shock remain at a high temperature, and the lack of ejecta emission relative to CO is not due to cooling of ejecta.

%Figure7

\begin{figure*}
\includegraphics[scale=0.6,angle=0,width=15truecm]{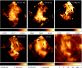}
%{Figures/IFUnorthcaivCO6panels2.png}
\caption{Cas A band images of the northern IFU field: (a) [Ca~IV] at 3.2 \mic, (b) [Si~IX] at 3.9 \mic, (c) [Mg~IV] at 4.5\mic, (d) [Ar~VI] at 4.52 \mic, (e) CO 4.79 (4.788-4.798) \mic, and (f) synchrotron subtracted CO broad band (from F444W) images. The latter captures the brightest CO band. The CO emission appears similar to other ejecta maps.}
\label{casabandmaps}
\end{figure*}

\begin{figure*}
\begin{center}
\includegraphics[scale=0.6,angle=0,width=5.5truecm]{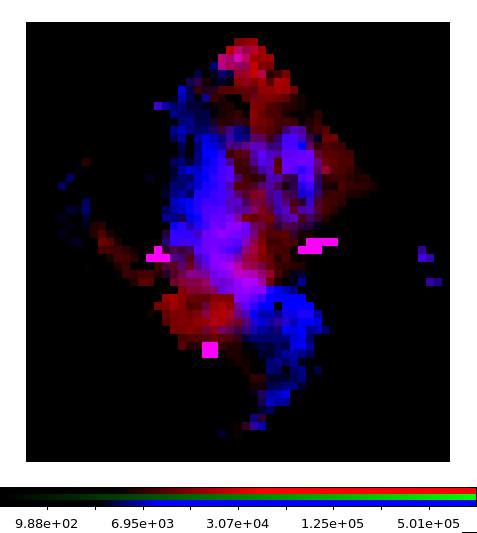}
%{Figures/caiv3p2umblueandredshiftmaps.png}
\includegraphics[scale=0.6,angle=0,width=5.5truecm]{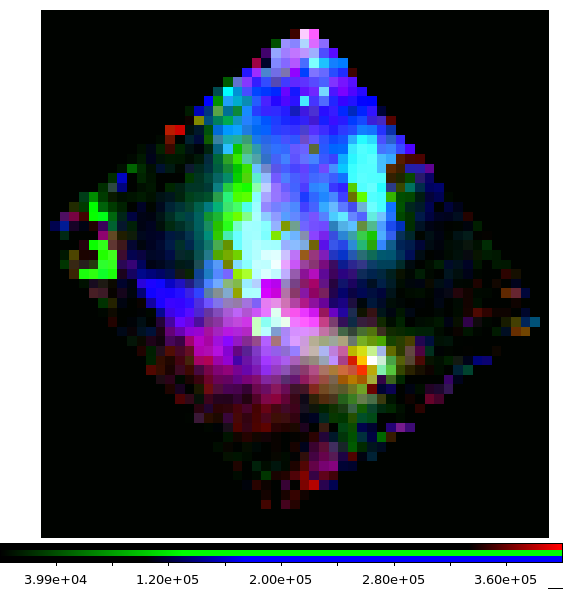}
%{Figures/casaIFUmapbMgIVgCOrSiIX.png}
\includegraphics[scale=0.6,angle=0,width=5.5truecm]{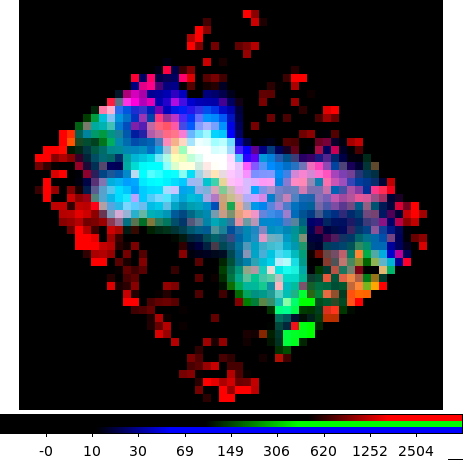}
%{Figures/IFUsouthrSiIXgCObMgIVnosmo.png}
\caption{Cas A band images of the northern IFU field: (Left) Red and blue colors denote the two different kinematic components of [Ca~IV] 3.2 \mic\ in this field. (Middle) Three color images showing [Mg~IV] (in blue), CO (green) and [Si~IX] (red). (Right) Three color images of the southern IFU field: [Mg~IV] (in blue), CO (green) and [Si~IX] (red). CO and [Mg~IV] show similar morphology while [Si~IX] morphology is different.}
\label{casabandmapscolor}
\end{center}
\end{figure*}

We compare the morphology of 4.5 \mic\ (4.3-5.2\mic) CO images with the IFU images of [Ca~IV] at 3.2 \mic, [Si~IX] at 3.9 \mic, [Mg~IV] at 4.5 \mic, [Ar~VI] at 4.52 \mic. Figure \ref{casabandmaps} shows separate ejecta distributions of Ca, Si, Mg, and Ar emission compared with CO emission in the northern NIRSpec IFU position. The Mg map shows more emission in the northern part and different morphology than the others. The Si originates interior to the layers than C, O, or Mg. Figure~\ref{casabandmapscolor} shows the spatial distribution of the two blue-shifted (see Table \ref{Tlinefluxes}) Mg emission components, color coded as blue and red in the Figure; the blue-colored component stretched north to south, and the red-colored component stretches from southeast to northwest. Three color maps of [Mg~IV], CO, and [Si~IX] emission in Figure \ref{casabandmapscolor} show contrasting distributions, as the Si map differs from the other two. Figure \ref{casabandmapscolor} shows the southern IFU position, where again the Si map looks different from the CO and Mg ejecta emission distributions. The contrasting distribution of [Si~IX] could be due to the fact that the [Si~IX] probably comes from a region of much lower density than the lower ionization lines, since the temperature is inversely proportional to the density ($n_0$) in a shock if the ram pressure is constant. More extensive IFU coverage of Cas A can help advance our understanding of the relative locations in ejecta, CO, and dust.

%Figure8

%Figure9

\subsection{CO modeling and properties}
\label{SLTE}

The Cas A spectra in Figure \ref{casanirspecCO} show two large rounded peaks corresponding to the R and P fundamental branches at 4.58 and 4.83 \mic, respectively \citep[see the discussion therein]{rho12}. The interval (wavelength difference) between R and P branch peaks depends on the temperature. The width between them is narrower for a lower temperature \citep{cami10,rho12}. The spectra show about 70 sawtooth-like CO spectral features in Figure \ref{casanirspecCO}. The contrast in the CO lines in the northern region spectra (red in Figure \ref{casanirspecCOorion}) is higher than that seen in the southern spectra (green). The CO spectral features largely correspond to specific v=1-0 vibrational transitions from various rotational levels, $J$. Intrinsic and instrumental broadening of the lines blends then with weaker lines from v=2-1 and higher vibrational states, and with each other for higher $J$ values in the R branch.

The CO fundamental lines have a similarity to the JWST spectra seen in the Orion nebula (blue in Figure \ref{casanirspecCOorion}), indicating similar individual CO features between Cas A and Orion. However, Cas A has much larger kinematic broadening which changes the spectral appearance. The wings of the lines, and weaker lines, blend together to creating a pseudo-continuum. Recent JWST observations of HH 211 detect CO features \citep{ray23}, somewhat similar to those of Cas A, but its pseudo-continuum is not apparent like in Cas A. The projected dynamic velocity in HH 211 is 80 - 100 \kms, which can be the upper limit of the CO line width, and the line width is likely much less than that in Cas A.

We performed LTE model fitting based on \cite{cami10} and the CO molecular line data from \cite{goorvitch94} to determine the physical conditions of CO molecules.  This approach is similar to that used by \cite{rho12, banerjee16, rho18sn, rho21}. We have quantified the goodness of fit ($\chi^2$) for the models using the IDL routine {\sc mpfit} and the python routines of {\sc mpfit} and {\sc lmfit}  for the best-fitting parameters \citep{markwardt09}. Before we compared with the CO LTE models, we shifted the velocity of the CO features to match the red- and blue-shifted fine-structure line of [Ar~VI] at 4.53 \mic\ (we used the brighter component among the two velocity components). The northern and southern knots are blue- and red-shifted, respectively. This shift allows the CO models to match the individual ro-vibrational lines.

%Figure \ref{casaCOLTE} shows an LTE model with a temperature of 1078 (900 - 1250) K. %The statistical errors are $\sim$2\%, but the errors include systematic errors by using different routines and methods. The intensities of the CO lines are well reproduced by the model, but the contrast in the lines is lower at higher $J$ in the R branch, and the level of the pseudo-continuum falls well below the data at high $J$ (the extreme long and short wavelengths of the emission). 

\begin{figure*}[!th]
\begin{center}
\includegraphics[scale=0.6,angle=0,width=8.5truecm,height=6.7truecm]
{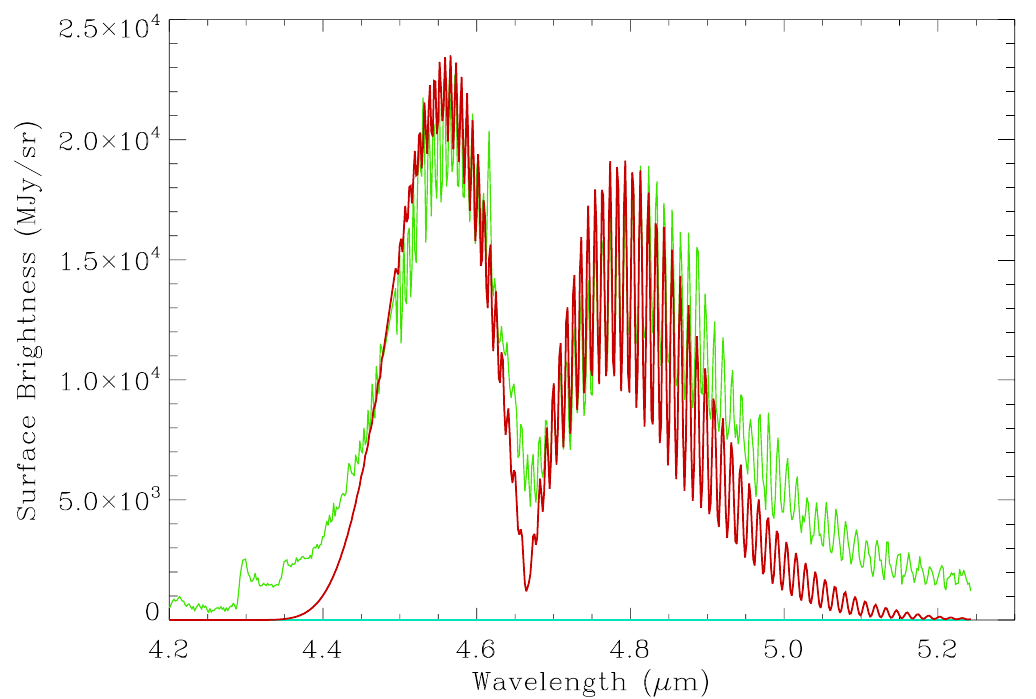}
\includegraphics[scale=0.6,angle=0,width=8.5truecm,height=6.7truecm]
{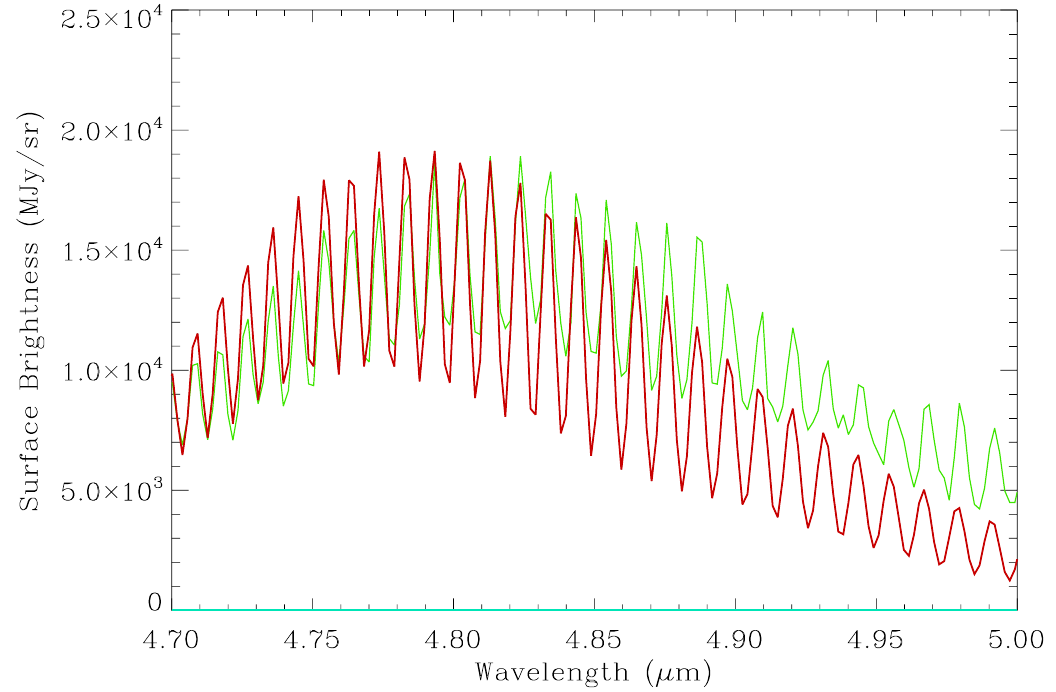}
\caption{An LTE model (in red) using a temperature ($T$) of 1078 K and a velocity width of 200 \kms\ is superposed on the JWST southern region spectrum of Cas A (in green). The sawtooth-like patterns are reproduced, but the relative strengths differ slightly. In the expanded view in the lower panel, the peaks are from CO v=1-0 P(4) through P(32) transitions. Blends with weaker v=2-1 and 
higher vibrational states modulate the apparent contrast of the lines.}
%The contribution of synchrotron emission is negligible ($< 1$\%).}
\label{casaCOLTE}
\end{center}
\end{figure*}

Figure~\ref{casaCOLTE} shows the best fit of an LTE model with a temperature of 1078 (900 - 1250) K, a velocity width (FWHM) of 237 (160 - 350) \kms, and a column density of 3.8 (1-20) $\times$10$^{17}$ cm$^{-2}$.
%for which the flux density unit was scaled to match to the JWST image of CO map from the NIRCam image F444W.}
The statistical errors are $\sim$2\%, but the errors include systematic errors by using different routines and methods. 
The intensities of the CO lines are well reproduced by the model, but the contrast in the lines is lower at higher $J$ in the R branch, and the level of the pseudo-continuum falls well below the data at high $J$ (the extreme long and short wavelengths of the emission). 
The spectral resolving power of the IFU is $\sim1000$ at these wavelengths, which translates to a velocity resolution of $\sim111$ \kms. However wider Gaussian profiles with a velocity width of 237 (160 - 350) \kms\ were able to reproduce the observed contrast of the CO lines and the blended emission underneath, as shown in Figure \ref{casaCOLTE}. Isolated far-IR CO rotational lines have shown similar velocity widths \citep{wallstrom13}. After we used linewidth broadening, no extra smoothing was needed. The fundamental band is a combination of hundreds of individual lines, while the far-IR CO rotational lines are well isolated at each wavelength. The best-fit model (red in Figure \ref{casaCOLTE}) with a velocity width of 237 \kms\ has too little contrast in the lines compared to the observed spectrum (green in Figure \ref{casaCOLTE}) in the R branch CO features ($<$ 4.67 \mic), which suggests that a lower velocity width may be appropriate.  

The ejecta and molecular knots have different density regions within them. The velocity width of $\sim$237 \kms\ indicates that the CO gas is the post-shocked region of the reverse shock.
The dense knots are compressed and fragmented by instabilities arising from the shock passage \citep{mellema02, raga07, silvia10}. The turbulence in dense knots is small, less than the turbulent velocity of 10 \kms\ in the warm gas \citep{korpi99}. 

The observed velocity width of CO is the internal shock velocity  that results from varying CO density and geometry (including the scales unresolved) 
that lead to variation in the shock speed and the direction with respect to the line of sight.
The shock velocity in CO gas ($V_{\rm CO}$) is inversely proportional to the square root of the density ratio between CO ($n_{\rm CO}$) and ejecta 
($n_{\rm ejecta}$) and proportional to the reverse shock velocity ($V_{\rm rs}$) ($V_{\rm CO}$= $V_{\rm rs}\times\sqrt{(n_{\rm ejecta}/n_{\rm CO}}$) owing to the conservation of energy \citep{biscaro14, klein03}. 
The reverse shock velocity $V_{\rm rs}$ relative to the ejecta has been estimated to be 2000 - 3000 \kms\ using optical data \citep{morse04, biscaro14}.
%which is the expanding velocity subtracted by the proper motion velocity \citep[see equation 1 in][]{morse04}.} 

The reverse shock velocity in the ejecta can also be estimated from an X-ray gas temperature of 2.6 keV \citep[$\sim$3$\times$10$^7$ K;][]{hwang04}.
The X-ray gas temperature $T_{\rm x}$ \citep [= $3\over{16}$$~\mu\over{k}$~$V_{\rm rs}^2$ $\sim$ 11\,$V_{\rm rs}^2$; references in][]{rho21} infers V$_{rs}$ of 1700 \kms\ \citep[Equation 12-24 in][]{spitzer78} where $\mu$ is the mean mass per particle and k is the Boltzmann constant). The density of the ejecta is 10$^{4-5}$ cm$^{-3}$ \citep{deLaney10}, and the density of CO gas is approximately 10$^{6-7}$ cm$^{-3}$ \citep{biscaro14,docenko10}. 
The critical density of CO  ($n_{\rm crit}$ =  A$_{ul}$/$\gamma_{ul}$ 
where the Einstein A is the
radiative transition, and $\gamma$ is the collision coefficient between upper and lower levels) is high ($>$ 10$^6$ cm$^{-3}$) for a temperature of 1000 K and an example of the upper and lower levels of v = 21 $\rightarrow$ 20\footnote{\url{https://home.strw.leidenuniv.nl/~moldata/datafiles/co.dat}}. 
The observed velocity width of 237 (160 - 350) \kms\ is approximately the shock velocity inside the CO clumps of
170 - 300 \kms\ ( = 1700 - 3000 $\times~\sqrt{(5\times\,10^{4})/(5\times\,10^6)}$.
% ($\sim$ 200 - 300 \kms).}

The LTE model to reproduce the JWST spectra of Cas A only uses $^{12}$C$^{16}$O without other CO isotopes. We simulated the intensity profile for optically thin and thick cases and found that the CO emission in Cas A was close to the optically thin case as it was found in the low-resolution AKARI spectra \citep{rho12}. The CO column density of 3.8 (1 - 20) $\times$10$^{17}$ cm$^{-2}$ is comparable to those obtained from CO excitation diagram \citep{wallstrom13}. The high column density is consistent with the idea that CO molecules are reformed behind the reverse shock \citep{wallstrom13, biscaro14}. The corresponding CO mass of the knot in the southern field is $\sim$2.3 $\times$10$^{-7}$ M$_\odot$ per pixel (0.1$''$)  at a distance of 3.4 kpc. Eventually, summing up all CO emissions from F444W emission and estimating the total CO mass from Cas A will be an interesting value. For that, JWST's advanced pipeline processing of individual frames is required, including removing hot pixels due to cosmic rays and various artifacts related to star subtraction, which is out of the scope of this paper. The dip between the R and P branches (around 4.65 \mic) from the LTE model is slightly off from the JWST spectra that were shifted based on ejecta lines. When we made a further shift of -0.014 \mic\ (a total of 0.03 \mic), the dip in the model was consistent with the JWST spectra of Cas A.

\begin{figure}
\begin{center}
\includegraphics[scale=0.6,angle=0,width=8.5truecm,height=7.truecm]{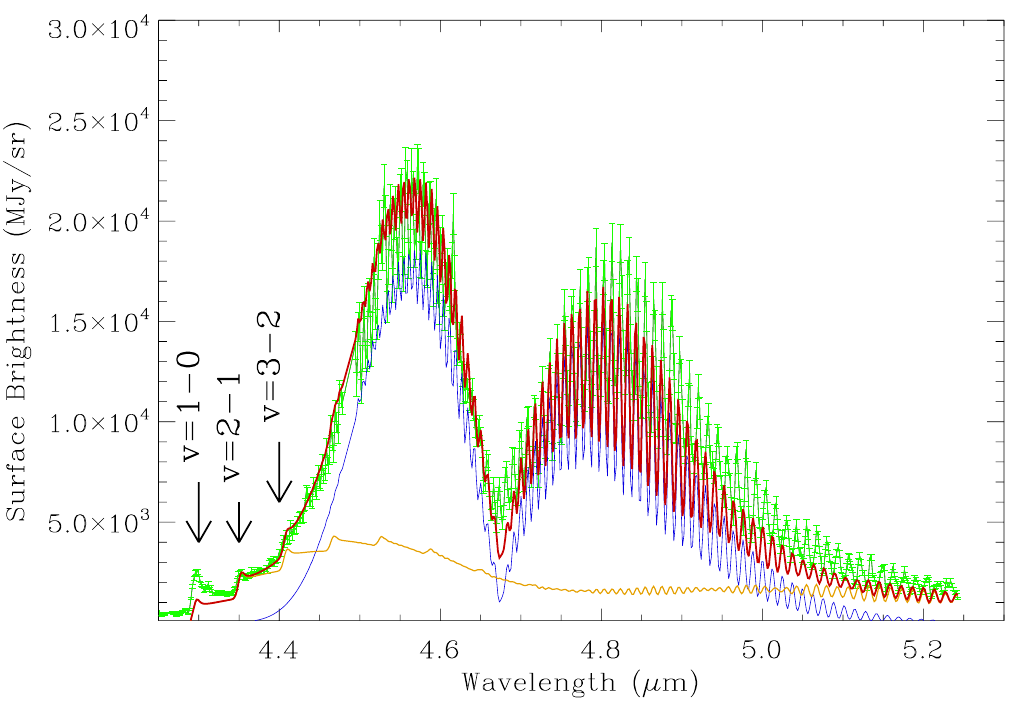}
%{Figures/COtwotemp3.pdf}
\caption{A two-temperature LTE CO model (in red) is superposed on the JWST southern spectrum of Cas A (in green). Here we have added a hotter CO component with $T = 4800$ K (yellow) and $\Delta V = 210$ \kms\ to the first temperature component (blue). The total fit is shown in red. The band heads (v=1-0, v=2-1, and v=3-2) are marked. The contribution of synchrotron emission is negligible about $\sim$50 MJy sr$^{-1}$ and wavelength dependence is small (the slops is 0.67).}
\label{casaCOLTE2}
\end{center}
\end{figure}

We notice a difference between the CO model and the JWST spectra at shorter and longer wavelengths as shown in Figures \ref{casaCOLTE}  and \ref{casaCOLTE2}, which is much larger than the synchrotron emission ($\sim$5 MJy sr$^{-1}$). The synchrotron emission is an interpolation from radio fluxes,  and detailed descriptions can be found in \cite{rho03, rho08, rho12}. The emission could be due to dust emission. Therefore, we added a continuum component using carbon dust to the CO component and fit the spectrum. The results show the continuum at 4.8-5.2 \mic\ has improved (the reduced $\chi^2$ improved a factor 10), but the continuum fit did not improve the residuals between 4.25 and 4.4 \mic.
%Figure10

%Figure11

An interesting feature of the Cas A CO spectra is the band heads that appear at $\sim4.3$ \mic. These band heads are a rare detection. Starting from $J=0$ at $\sim 4.6$ \mic, the wavelengths of successive R branch transitions get progressively shorter, as $J$ increases. However, the lines become increasingly closely spaced, and after $J\gtrsim 90$ the lines begin increasing in wavelength. The unresolvable pile up in wavelength of lines near $J\sim 90$ creates the band heads. Cas A spectra clearly show these band heads for at least three different vibrational transitions (v=1-0 at 4.3 \mic, v=2-1 at 4.35 \mic, and v=3-2 at 4.45 \mic), marked in Figure \ref{casanirspecregions}e. The laboratory band head wavelengths are v=1-0 at 4.295 \mic, v=2-1 at 4.350 \mic, v= 3-2 at 4.295 \mic, and v= 4-3 at 4.465 \mic\ \citep{geballe07}. The observed band heads are shifted by about -2000 \kms\ and +1000 \kms\ for the northern and southern region spectra, respectively, as we have previously noted. The excitation energy of the v=0 J=90 level is about $\sim 22,000$ K, which implies that the temperature must be high for even a small fraction of the CO to be excited to such high v and J states. Therefore, the presence of such band heads may indicate multi-temperature CO gas, or it may require a non-LTE model. We attempted to fit a two-temperature CO LTE model (with a second temperature component of 4800 K), shown in Figure \ref{casaCOLTE2}, although the fit did not converge efficiently due to many small sinusoidal patterns and diverse combinations of two CO models. However, it demonstrates that the high-temperature CO model can reproduce the CO band heads at 4.3, 4.35, and 4.45 \mic, indicating multiple temperature components of CO gas likely exist.

We also analyzed the CO excitation with the non-LTE radiative transfer code  {\sc RADEX} \citep{vanderTak07} as it was done by \cite{wallstrom13}. The more sophisticated methods retain the assumption of a local excitation, but solve for the balance of excitation and de-excitation rates from and to a given state, the so-called statistical equilibrium (SE). The {\sc RADEX} program was adapted for MIR and NIR CO emission \citep{wallstrom13}. While the model reproduced low-resolution CO AKARI spectra using temperatures of 400 and 2000 K, a velocity width of 300 \kms, and $N_{\rm CO}$ of 4$\times$10$^{17}$ cm$^{-2}$, it could not reproduce the CO sawtooth-like patterns seen in the JWST spectra. The contrast in the CO lines was much larger than in the JWST spectra (more than a factor of 5 stronger) when using a velocity of $\sim$200-600 \kms, and increasing the velocity did not help reproduce the JWST spectra. The {\sc RADEX} model produces asymmetrically oscillating CO patterns, which differ from the observed JWST spectra. The model, in particular, lacks CO emission below 4.5 \mic\ and above 4.9 \mic, which could be due to lack of H-atom collision partners or other missing collision rates. Future work on non-LTE modeling is encouraged to reproduce the JWST spectra of Cas A, which may need to include a large Doppler velocity broadening  and various collision rates, including H-atom collision partners.

Non-LTE CO modeling can be preformed using the HYDrodynamical RAdiation code (HYDRA), a hydrodynamic SN evolution code, including the nuclear networks and ejecta mixing, and Radiation code, accounting for the atomic level population and the equations of state \citep[see][]{hoeflich93, rho21}. The $\gamma$-rays and free electrons at a certain temperature can ionize CO, which may affect the timescale of the CO and CO$^+$ formation and the ratio of CO$^+$ and CO. The time-dependent rate equations are  $C + O \rightarrow CO$ and $C^+ + O \rightarrow CO^+$ \citep[see also][]{liljegren23,ono24}. Nonthermal excitation can enhance CO$^+$, which adds a blue-shifted emission component. NIR and MIR spectral signatures from CO only and from CO and CO$^+$ with and without non-thermal excitation for SNe Ic are described in the Appendix of \cite{rho21}. Such complex HYDRA modeling for the progenitor SN IIb is encouraged and can be applied to the JWST spectra of Cas A.

\vskip -2truecm
\section*{Conclusion}

In this paper we have presented and analyzed JWST observations of the supernova remnant Cas A to investigate the physical conditions for molecular CO formation and destruction in supernova ejecta. Our conclusions can be summarized as follows:

1. NIRCam (F356W and F444W) and MIRI (F770W) images map synchrotron radiation, Ar-rich ejecta, and  CO in great detail and reveal complex structures on the scale of individual ejecta knots. The CO map strongly correlates with knotty ejecta structures. CO is brightest in the northern and southern portions of the main shell, and weakest in the east where ejecta emission dominates.

2. Complex fundamental bands of CO features dominate the NIRSpec-IFU spectra, with $\sim 70$ individually resolvable ro-vibrational lines. These lines are similar to those seen in the Orion nebula, but with instrumental and velocity broadening in Cas A that blends lines to produce a pseudo-continuum. 

3. LTE modeling of the CO spectra (R and P branches) suggests an optically thin CO gas with a temperature of $\sim1080$ K, $N({\rm CO}) = 4 \times 10^{17}$\,cm$^{-2}$, $v$=200 \kms, and $M_{\rm CO}$ = 2$\times$10$^{-7}$ $M_\odot$ per 0.1\arcsec\ pixel.

4. We discover band heads for the v=1-0, 2-1, and 3-2 CO vibrational transitions at 4.3, 4.35, and 4.4 \mic. The high rotational levels ($J\sim 90$) that need to be populated to form the band head emission requires a high temperature component in modeling the emission. When we fit the spectrum including a second CO component, the second temperature is $\sim$4800 K, and the fitting favors a lower first temperature component (i.e., lower than 1080 K), indicating the presence of multiple temperatures.\\

\vskip 0.4truecm
%\begin{acknowledgements}
%\acknowledgement
We thank the anonymous referee for the insightful comments. We thank Jan Cami, John Black,  Sofia Wallstrom, and F. van der Tak for their assistance and discussion of CO LTE and non-LTE models. We thank Tom Geballe for the fruitful discussion regarding the various line transitions of CO molecules. This work is based on observations made with the NASA/ESA/CSA James Webb Space Telescope. Support for this program \#1947 was provided by NASA through a grant from the Space Telescope Science Institute, which is operated by the Association of Universities for Research in Astronomy, Inc., under NASA contract NAS 5-03127. J.R. was supported by JWST-GO-01947.032, and was partially supported by a NASA ADAP grant (80NSSC23K0749) and Brain Pool visiting program for Outstanding Overseas Researchers by the National Research Foundation of Korea (NRF-2022H1D3A2A01096434). S.-H. P. and S.-C.Y. was supported by NRF-2019R1A2C2010885 and NRF-2022H1D3A2A01096434. R.G.A. was supported by NASA under award number 80GSFC21M0002. J.M.L.\ was supported by JWST grant JWST-GO-01947.023 and by basic research funds of the Office of Naval Research. H.-T. J. is grateful for support from the German Research Foundation (DFG) through the Collaborative Research Centre ``Neutrinos and Dark Matter in Astro- and Particle Physics (NDM),'' Grant No.\ SFB-1258-283604770, and under Germany's Excellence Strategy through the Cluster of Excellence ORIGINS EXC-2094-390783311.  
%\end{acknowledgements}

\bibliography{msrefsall}

\begin{thebibliography}{}
\expandafter\ifx\csname natexlab\endcsname\relax\def\natexlab#1{#1}\fi

\bibitem[{{Anderson} {et~al.}(1991){Anderson}, {Rudnick}, {Leppik}, {Perley},
  \& {Braun}}]{anderson91}
{Anderson}, M., {Rudnick}, L., {Leppik}, P., {Perley}, R., \& {Braun}, R. 1991,
  \apj, 373, 146

\bibitem[{{Banerjee} {et~al.}(2016){Banerjee}, {Srivastava}, {Ashok}, \&
  {Venkataraman}}]{banerjee16}
{Banerjee}, D.~P.~K., {Srivastava}, M.~K., {Ashok}, N.~M., \& {Venkataraman},
  V. 2016, \mnras, 455, L109

\bibitem[{{Biscaro} \& {Cherchneff}(2014)}]{biscaro14}
{Biscaro}, C., \& {Cherchneff}, I. 2014, \aap, 564, A25

\bibitem[{{Borkowski} \& {Shull}(1990)}]{borkowski90}
{Borkowski}, K.~J., \& {Shull}, J.~M. 1990, \apj, 348, 169

\bibitem[{{Cami} {et~al.}(2010){Cami}, {Bernard-Salas}, {Peeters}, \&
  {Malek}}]{cami10}
{Cami}, J., {Bernard-Salas}, J., {Peeters}, E., \& {Malek}, S.~E. 2010,
  Science, 329, 1180

\bibitem[{{Chawner} {et~al.}(2019){Chawner}, {Marsh}, {Matsuura}, {Gomez},
  {Cigan}, {De Looze}, {Barlow}, {Dunne}, {Noriega-Crespo}, \&
  {Rho}}]{chawner19}
{Chawner}, H., {Marsh}, K., {Matsuura}, M., {et~al.} 2019, \mnras, 483, 70

\bibitem[{{Cherchneff} \& {Dwek}(2009)}]{cherchneff09}
{Cherchneff}, I., \& {Dwek}, E. 2009, \apj, 703, 642

\bibitem[{{Cherchneff} \& {Lilly}(2008)}]{cherchneff08}
{Cherchneff}, I., \& {Lilly}, S. 2008, \apjl, 683, L123

\bibitem[{{Chevalier}(1977)}]{chevalier77}
{Chevalier}, R.~A. 1977, \araa, 15, 175

\bibitem[{{De Looze} {et~al.}(2017){De Looze}, {Barlow}, {Swinyard}, {Rho},
  {Gomez}, {Matsuura}, \& {Wesson}}]{delooze17}
{De Looze}, I., {Barlow}, M.~J., {Swinyard}, B.~M., {et~al.} 2017, \mnras, 465,
  3309

\bibitem[{{De Looze, I.}(2024)}]{deLooze24}
{De Looze, I.} 2024, ApJL, in prep.

\bibitem[{{DeLaney} {et~al.}(2014){DeLaney}, {Kassim}, {Rudnick}, \&
  {Perley}}]{deLaney14}
{DeLaney}, T., {Kassim}, N.~E., {Rudnick}, L., \& {Perley}, R.~A. 2014, \apj,
  785, 7

\bibitem[{{DeLaney} {et~al.}(2010){DeLaney}, {Rudnick}, {Stage}, {Smith},
  {Isensee}, {Rho}, {Allen}, {Gomez}, {Kozasa}, {Reach}, {Davis}, \&
  {Houck}}]{deLaney10}
{DeLaney}, T., {Rudnick}, L., {Stage}, M.~D., {et~al.} 2010, \apj, 725, 2038

\bibitem[{{Docenko} \& {Sunyaev}(2010)}]{docenko10}
{Docenko}, D., \& {Sunyaev}, R.~A. 2010, \aap, 509, A59 (DS10)

\bibitem[{{Dwek} \& {Cherchneff}(2011)}]{dwek11}
{Dwek}, E., \& {Cherchneff}, I. 2011, \apj, 727, 63

\bibitem[{{Dwek} {et~al.}(2019){Dwek}, {Sarangi}, \& {Arendt}}]{dwek19}
{Dwek}, E., {Sarangi}, A., \& {Arendt}, R.~G. 2019, \apjl, 871, L33

\bibitem[{{Ennis} {et~al.}(2006){Ennis}, {Rudnick}, {Reach}, {Smith}, {Rho},
  {DeLaney}, {Gomez}, \& {Kozasa}}]{ennis06}
{Ennis}, J.~A., {Rudnick}, L., {Reach}, W.~T., {et~al.} 2006, \apj, 652, 376

\bibitem[{{Fazio} {et~al.}(2004){Fazio}, {Hora}, {Allen}, {Ashby}, {Barmby},
  {Deutsch}, {Huang}, {Kleiner}, {Marengo}, {Megeath}, {Melnick}, {Pahre},
  {Patten}, {Polizotti}, {Smith}, {Taylor}, {Wang}, {Willner}, {Hoffmann},
  {Pipher}, {Forrest}, {McMurty}, {McCreight}, {McKelvey}, {McMurray}, {Koch},
  {Moseley}, {Arendt}, {Mentzell}, {Marx}, {Losch}, {Mayman}, {Eichhorn},
  {Krebs}, {Jhabvala}, {Gezari}, {Fixsen}, {Flores}, {Shakoorzadeh}, {Jungo},
  {Hakun}, {Workman}, {Karpati}, {Kichak}, {Whitley}, {Mann}, {Tollestrup},
  {Eisenhardt}, {Stern}, {Gorjian}, {Bhattacharya}, {Carey}, {Nelson},
  {Glaccum}, {Lacy}, {Lowrance}, {Laine}, {Reach}, {Stauffer}, {Surace},
  {Wilson}, {Wright}, {Hoffman}, {Domingo}, \& {Cohen}}]{fazio04}
{Fazio}, G.~G., {Hora}, J.~L., {Allen}, L.~E., {et~al.} 2004, \apjs, 154, 10

\bibitem[{{Fesen}(2001)}]{fesen01}
{Fesen}, R.~A. 2001, \apjs, 133, 161

\bibitem[{{Fesen} {et~al.}(2006){Fesen}, {Hammell}, {Morse}, {Chevalier},
  {Borkowski}, {Dopita}, {Gerardy}, {Lawrence}, {Raymond}, \& {van den
  Bergh}}]{fesen06}
{Fesen}, R.~A., {Hammell}, M.~C., {Morse}, J., {et~al.} 2006, \apj, 645, 283

\bibitem[{{Gall} {et~al.}(2011){Gall}, {Hjorth}, \& {Andersen}}]{gall11}
{Gall}, C., {Hjorth}, J., \& {Andersen}, A.~C. 2011, \aapr, 19, 43

\bibitem[{{Geballe} {et~al.}(2007){Geballe}, {Rushton}, {Eyres}, {Evans}, {van
  Loon}, \& {Smalley}}]{geballe07}
{Geballe}, T.~R., {Rushton}, M.~T., {Eyres}, S.~P.~S., {et~al.} 2007, \aap,
  467, 269

\bibitem[{{Goorvitch}(1994)}]{goorvitch94}
{Goorvitch}, D. 1994, \apjs, 95, 535

\bibitem[{{Hoflich} {et~al.}(1993){Hoflich}, {Langer}, \&
  {Duschinger}}]{hoeflich93}
{Hoflich}, P., {Langer}, N., \& {Duschinger}, M. 1993, \aap, 275, L29

\bibitem[{{Hwang} {et~al.}(2004){Hwang}, {Laming}, {Badenes}, {Berendse},
  {Blondin}, {Cioffi}, {DeLaney}, {Dewey}, {Fesen}, {Flanagan}, {Fryer},
  {Ghavamian}, {Hughes}, {Morse}, {Plucinsky}, {Petre}, {Pohl}, {Rudnick},
  {Sankrit}, {Slane}, {Smith}, {Vink}, \& {Warren}}]{hwang04}
{Hwang}, U., {Laming}, J.~M., {Badenes}, C., {et~al.} 2004, \apjl, 615, L117

\bibitem[{{Isensee} {et~al.}(2010){Isensee}, {Rudnick}, {DeLaney}, {Smith},
  {Rho}, {Reach}, {Kozasa}, \& {Gomez}}]{Isensee10}
{Isensee}, K., {Rudnick}, L., {DeLaney}, T., {et~al.} 2010, \apj, 725, 2059

\bibitem[{{Isensee} {et~al.}(2012){Isensee}, {Olmschenk}, {Rudnick}, {DeLaney},
  {Rho}, {Smith}, {Reach}, {Kozasa}, \& {Gomez}}]{Isensee12}
{Isensee}, K., {Olmschenk}, G., {Rudnick}, L., {et~al.} 2012, \apj, 757, 126

\bibitem[{{Kirchschlager} {et~al.}(2023){Kirchschlager}, {Schmidt}, {Barlow},
  {De Looze}, \& {Sartorio}}]{kirchschlager23}
{Kirchschlager}, F., {Schmidt}, F.~D., {Barlow}, M.~J., {De Looze}, I., \&
  {Sartorio}, N.~S. 2023, \mnras, 520, 5042

\bibitem[{{Kirchschlager} {et~al.}(2019){Kirchschlager}, {Schmidt}, {Barlow},
  {Fogerty}, {Bevan}, \& {Priestley}}]{kirchschlager19}
{Kirchschlager}, F., {Schmidt}, F.~D., {Barlow}, M.~J., {et~al.} 2019, \mnras,
  489, 4465

\bibitem[{{Klein} {et~al.}(2003){Klein}, {Budil}, {Perry}, \& {Bach}}]{klein03}
{Klein}, R.~I., {Budil}, K.~S., {Perry}, T.~S., \& {Bach}, D.~R. 2003, \apj,
  583, 245

\bibitem[{{Koo} {et~al.}(2020){Koo}, {Kim}, {Oh}, {Raymond}, {Yoon}, {Lee}, \&
  {Jaffe}}]{koo20}
{Koo}, B.-C., {Kim}, H.-J., {Oh}, H., {et~al.} 2020, Nature Astronomy, 4, 584

\bibitem[{{Korpi} {et~al.}(1999){Korpi}, {Brandenburg}, {Shukurov}, \&
  {Tuominen}}]{korpi99}
{Korpi}, M.~J., {Brandenburg}, A., {Shukurov}, A., \& {Tuominen}, I. 1999,
  \aap, 350, 230

\bibitem[{{Kotak} {et~al.}(2009){Kotak}, {Meikle}, {Farrah}, {Gerardy},
  {Foley}, {Van Dyk}, {Fransson}, {Lundqvist}, {Sollerman}, {Fesen},
  {Filippenko}, {Mattila}, {Silverman}, {Andersen}, {H{\"o}flich}, {Pozzo}, \&
  {Wheeler}}]{kotak09}
{Kotak}, R., {Meikle}, W.~P.~S., {Farrah}, D., {et~al.} 2009, \apj, 704, 306

\bibitem[{{Krause} {et~al.}(2008){Krause}, {Birkmann}, {Usuda}, {Hattori},
  {Goto}, {Rieke}, \& {Misselt}}]{krause08}
{Krause}, O., {Birkmann}, S.~M., {Usuda}, T., {et~al.} 2008, Science, 320, 1195

\bibitem[{{Laporte} {et~al.}(2017){Laporte}, {Ellis}, {Boone}, {Bauer},
  {Qu{\'e}nard}, {Roberts-Borsani}, {Pell{\'o}}, {P{\'e}rez-Fournon}, \&
  {Streblyanska}}]{laporte17}
{Laporte}, N., {Ellis}, R.~S., {Boone}, F., {et~al.} 2017, \apjl, 837, L21

\bibitem[{{Liljegren} {et~al.}(2023){Liljegren}, {Jerkstrand}, {Barklem},
  {Nyman}, {Brady}, \& {Yurchenko}}]{liljegren23}
{Liljegren}, S., {Jerkstrand}, A., {Barklem}, P.~S., {et~al.} 2023, \aap, 674,
  A184

\bibitem[{{Markwardt}(2009)}]{markwardt09}
{Markwardt}, C.~B. 2009, in Astronomical Society of the Pacific Conference
  Series, Vol. 411, Astronomical Data Analysis Software and Systems XVIII, ed.
  D.~A. {Bohlender}, D.~{Durand}, \& P.~{Dowler}, 251

\bibitem[{{Matsuura} {et~al.}(2011){Matsuura}, {Dwek}, {Meixner}, {Otsuka},
  {Babler}, {Barlow}, {Roman-Duval}, {Engelbracht}, {Sandstrom},
  {Laki{\'c}evi{\'c}}, {van Loon}, {Sonneborn}, {Clayton}, {Long}, {Lundqvist},
  {Nozawa}, {Gordon}, {Hony}, {Panuzzo}, {Okumura}, {Misselt}, {Montiel}, \&
  {Sauvage}}]{matsuura11}
{Matsuura}, M., {Dwek}, E., {Meixner}, M., {et~al.} 2011, Science, 333, 1258

\bibitem[{{Matsuura} {et~al.}(2015){Matsuura}, {Dwek}, {Barlow}, {Babler},
  {Baes}, {Meixner}, {Cernicharo}, {Clayton}, {Dunne}, {Fransson}, {Fritz},
  {Gear}, {Gomez}, {Groenewegen}, {Indebetouw}, {Ivison}, {Jerkstrand},
  {Lebouteiller}, {Lim}, {Lundqvist}, {Pearson}, {Roman-Duval}, {Royer},
  {Staveley-Smith}, {Swinyard}, {van Hoof}, {van Loon}, {Verstappen}, {Wesson},
  {Zanardo}, {Blommaert}, {Decin}, {Reach}, {Sonneborn}, {Van de Steene}, \&
  {Yates}}]{matsuura15}
{Matsuura}, M., {Dwek}, E., {Barlow}, M.~J., {et~al.} 2015, \apj, 800, 50

\bibitem[{{Mellema} {et~al.}(2002){Mellema}, {Kurk}, \&
  {R{\"o}ttgering}}]{mellema02}
{Mellema}, G., {Kurk}, J.~D., \& {R{\"o}ttgering}, H.~J.~A. 2002, \aap, 395,
  L13

\bibitem[{{Milisavljevic} \& {Fesen}(2013)}]{milisavljevic13}
{Milisavljevic}, D., \& {Fesen}, R.~A. 2013, \apj, 772, 134

\bibitem[{{Milisavljevic} {et~al.}(2024){Milisavljevic}, {Temim}, {De Looze},
  {Dickinson}, {Laming}, {Fesen}, {Raymond}, {Arendt}, {Vink}, {Posselt},
  {Pavlov}, {Fox}, {Pinarski}, {Subrayan}, {Schmidt}, {Blair}, {Rest},
  {Patnaude}, {Koo}, {Rho}, {Orlando}, {Janka}, {Andrews}, {Barlow}, {Burrows},
  {Chevalier}, {Clayton}, {Fransson}, {Fryer}, {Gomez}, {Kirchschlager}, {Lee},
  {Matsuura}, {Niculescu-Duvaz}, {Pierel}, {Plucinsky}, {Priestley}, {Ravi},
  {Sartorio}, {Schmidt}, {Shahbandeh}, {Slane}, {Smith}, {Sravan}, {Weil},
  {Wesson}, \& {Wheeler}}]{milisavljevic24}
{Milisavljevic}, D., {Temim}, T., {De Looze}, I., {et~al.} 2024, \apjl, 965,
  L27

\bibitem[{{Millard} {et~al.}(2021){Millard}, {Ravi}, {Rho}, \&
  {Park}}]{millard21}
{Millard}, M.~J., {Ravi}, A.~P., {Rho}, J., \& {Park}, S. 2021, \apjs, 257, 36

\bibitem[{{Morse} {et~al.}(2004){Morse}, {Fesen}, {Chevalier}, {Borkowski},
  {Gerardy}, {Lawrence}, \& {van den Bergh}}]{morse04}
{Morse}, J.~A., {Fesen}, R.~A., {Chevalier}, R.~A., {et~al.} 2004, \apj, 614,
  727

\bibitem[{{Niculescu-Duvaz} {et~al.}(2022){Niculescu-Duvaz}, {Barlow}, {Bevan},
  {Wesson}, {Milisavljevic}, {De Looze}, {Clayton}, {Krafton}, {Matsuura}, \&
  {Brady}}]{niculescu-Duvaz22}
{Niculescu-Duvaz}, M., {Barlow}, M.~J., {Bevan}, A., {et~al.} 2022, \mnras,
  515, 4302

\bibitem[{{Nozawa} {et~al.}(2006){Nozawa}, {Kozasa}, \& {Habe}}]{nozawa06}
{Nozawa}, T., {Kozasa}, T., \& {Habe}, A. 2006, \apj, 648, 435

\bibitem[{{Nozawa} {et~al.}(2007){Nozawa}, {Kozasa}, {Habe}, {Dwek}, {Umeda},
  {Tominaga}, {Maeda}, \& {Nomoto}}]{nozawa07}
{Nozawa}, T., {Kozasa}, T., {Habe}, A., {et~al.} 2007, \apj, 666, 955

\bibitem[{{Nozawa} {et~al.}(2003){Nozawa}, {Kozasa}, {Umeda}, {Maeda}, \&
  {Nomoto}}]{nozawa03}
{Nozawa}, T., {Kozasa}, T., {Umeda}, H., {Maeda}, K., \& {Nomoto}, K. 2003,
  \apj, 598, 785

\bibitem[{{Ono} {et~al.}(2024){Ono}, {Nozawa}, {Nagataki}, {Kozyreva},
  {Orlando}, {Miceli}, \& {Chen}}]{ono24}
{Ono}, M., {Nozawa}, T., {Nagataki}, S., {et~al.} 2024, \apjs, 271, 33

\bibitem[{{Peeters} {et~al.}(2023){Peeters}, {Habart}, {Berne}, {Sidhu},
  {Chown}, {Van De Putte}, {Trahin}, {Schroetter}, {Canin}, {Alarcon},
  {Schefter}, {Khan}, {Pasquini}, {Tielens}, {Wolfire}, {Dartois},
  {Goicoechea}, {Maragkoudakis}, {Onaka}, {Pound}, {Vicente}, {Abergel},
  {Bergin}, {Bernard-Salas}, {Boersma}, {Bron}, {Cami}, {Cuadrado}, {Dicken},
  {Elyajour}, {Fuente}, {Gordon}, {Issa}, {Joblin}, {Kannavou}, {Lacinbala},
  {Languignon}, {Le Gal}, {Meshaka}, {Okada}, {Robberto}, {Roellig},
  {Schirmer}, {Tabone}, {Zannese}, {Aleman}, {Allamandola}, {Auchettl},
  {Baratta}, {Bejaoui}, {Bera}, {Black}, {Boulanger}, {Bouwman}, {Brandl},
  {Brechignac}, {Brunken}, {Buragohain}, {Burkhardt}, {Candian}, {Cazaux},
  {Cernicharo}, {Chabot}, {Chakraborty}, {Champion}, {Colgan}, {Cooke},
  {Coutens}, {Cox}, {Demyk}, {Donovan Meyer}, {Foschino}, {Garcia-Lario},
  {Gerin}, {Gottlieb}, {Guillard}, {Gusdorf}, {Hartigan}, {He}, {Herbst},
  {Hornekaer}, {Jager}, {Janot-Pacheco}, {Kaufman}, {Kendrew}, {Kirsanova},
  {Klaassen}, {Kwok}, {Labiano}, {Lai}, {Lee}, {Lefloch}, {Le Petit}, {Li},
  {Linz}, {Mackie}, {Madden}, {Mascetti}, {McGuire}, {Merino}, {Micelotta},
  {Misselt}, {Morse}, {Mulas}, {Neelamkodan}, {Ohsawa}, {Paladini}, {Palumbo},
  {Pathak}, {Pendleton}, {Petrignani}, {Pino}, {Puga}, {Rangwala}, {Rapacioli},
  {Ricca}, {Roman-Duval}, {Roser}, {Roueff}, {Rouille}, {Salama}, {Sales},
  {Sandstrom}, {Sarre}, {Sciamma-O'Brien}, {Sellgren}, {Shenoy}, {Teyssier},
  {Thomas}, {Togi}, {Verstraete}, {Witt}, {Wootten}, {Ysard}, {Zettergren},
  {Zhang}, {Zhang}, \& {Zhen}}]{peeters23}
{Peeters}, E., {Habart}, E., {Berne}, O., {et~al.} 2023, arXiv e-prints,
  arXiv:2310.08720

\bibitem[{{Priestley} {et~al.}(2021){Priestley}, {Chawner}, {Matsuura}, {De
  Looze}, {Barlow}, \& {Gomez}}]{priestley21}
{Priestley}, F.~D., {Chawner}, H., {Matsuura}, M., {et~al.} 2021, \mnras, 500,
  2543

\bibitem[{{Raga} {et~al.}(2007){Raga}, {Esquivel}, {Riera}, \&
  {Vel{\'a}zquez}}]{raga07}
{Raga}, A.~C., {Esquivel}, A., {Riera}, A., \& {Vel{\'a}zquez}, P.~F. 2007,
  \apj, 668, 310

\bibitem[{{Ray} {et~al.}(2023){Ray}, {McCaughrean}, {Caratti o Garatti},
  {Kavanagh}, {Justtanont}, {van Dishoeck}, {Reitsma}, {Beuther}, {Francis},
  {Gieser}, {Klaassen}, {Perotti}, {Tychoniec}, {van Gelder}, {Colina},
  {Greve}, {G{\"u}del}, {Henning}, {Lagage}, {{\"O}stlin}, {Vandenbussche},
  {Waelkens}, \& {Wright}}]{ray23}
{Ray}, T.~P., {McCaughrean}, M.~J., {Caratti o Garatti}, A., {et~al.} 2023,
  \nat, 622, 48

\bibitem[{{Rest} {et~al.}(2023){Rest}, {Pierel}, {Correnti}, {Canipe},
  {Hilbert}, {Engesser}, {Sunnquist}, \& {Fox}}]{rest23}
{Rest}, A., {Pierel}, J., {Correnti}, M., {et~al.} 2023, {arminrest/jhat: The
  JWST HST Alignment Tool (JHAT)}, Zenodo, doi:10.5281/zenodo.7892935

\bibitem[{{Rho} {et~al.}(2018{\natexlab{a}}){Rho}, {Geballe}, {Banerjee},
  {Dessart}, {Evans}, \& {Joshi}}]{rho18sn}
{Rho}, J., {Geballe}, T.~R., {Banerjee}, D.~P.~K., {et~al.} 2018{\natexlab{a}},
  \apjl, 864, L20

\bibitem[{{Rho} {et~al.}(2012){Rho}, {Onaka}, {Cami}, \& {Reach}}]{rho12}
{Rho}, J., {Onaka}, T., {Cami}, J., \& {Reach}, W.~T. 2012, \apjl, 747, L6

\bibitem[{{Rho} {et~al.}(2009){Rho}, {Reach}, {Tappe}, {Hwang}, {Slavin},
  {Kozasa}, \& {Dunne}}]{rho09}
{Rho}, J., {Reach}, W.~T., {Tappe}, A., {et~al.} 2009, \apj, 700, 579

\bibitem[{{Rho} {et~al.}(2003){Rho}, {Reynolds}, {Reach}, {Jarrett}, {Allen},
  \& {Wilson}}]{rho03}
{Rho}, J., {Reynolds}, S.~P., {Reach}, W.~T., {et~al.} 2003, \apj, 592, 299

\bibitem[{{Rho} {et~al.}(2008){Rho}, {Kozasa}, {Reach}, {Smith}, {Rudnick},
  {DeLaney}, {Ennis}, {Gomez}, \& {Tappe}}]{rho08}
{Rho}, J., {Kozasa}, T., {Reach}, W.~T., {et~al.} 2008, \apj, 673, 271

\bibitem[{{Rho} {et~al.}(2018{\natexlab{b}}){Rho}, {Gomez}, {Boogert}, {Smith},
  {Lagage}, {Dowell}, {Clark}, {Peeters}, \& {Cami}}]{rho18}
{Rho}, J., {Gomez}, H.~L., {Boogert}, A., {et~al.} 2018{\natexlab{b}}, \mnras,
  479, 5101

\bibitem[{{Rho} {et~al.}(2021){Rho}, {Evans}, {Geballe}, {Banerjee},
  {Hoeflich}, {Shahbandeh}, {Valenti}, {Yoon}, {Jin}, {Williamson}, {Modjaz},
  {Hiramatsu}, {Howell}, {Pellegrino}, {Vink{\'o}}, {Cartier}, {Burke},
  {McCully}, {An}, {Cha}, {Pritchard}, {Wang}, {Andrews}, {Galbany}, {Van Dyk},
  {Graham}, {Blinnikov}, {Joshi}, {P{\'a}l}, {Kriskovics}, {Ordasi}, {Szakats},
  {Vida}, {Chen}, {Li}, {Zhang}, \& {Yan}}]{rho21}
{Rho}, J., {Evans}, A., {Geballe}, T.~R., {et~al.} 2021, \apj, 908, 232

\bibitem[{{Rho} {et~al.}(2023){Rho}, {Ravi}, {Tram}, {Hoang}, {Chastenet},
  {Millard}, {Barlow}, {De Looze}, {Gomez}, {Kirchschlager}, \&
  {Dunne}}]{rho23casapol}
{Rho}, J., {Ravi}, A.~P., {Tram}, L.~N., {et~al.} 2023, \mnras, 522, 2279

\bibitem[{{Sarangi} \& {Cherchneff}(2013)}]{sarangi13}
{Sarangi}, A., \& {Cherchneff}, I. 2013, \apj, 776, 107

\bibitem[{{Sarangi} \& {Cherchneff}(2015)}]{sarangi15}
---. 2015, \aap, 575, A95

\bibitem[{{Sarangi} {et~al.}(2018){Sarangi}, {Matsuura}, \&
  {Micelotta}}]{sarangi18}
{Sarangi}, A., {Matsuura}, M., \& {Micelotta}, E.~R. 2018, \ssr, 214, 63

\bibitem[{{Shahbandeh} {et~al.}(2023){Shahbandeh}, {Sarangi}, {Temim},
  {Szalai}, {Fox}, {Tinyanont}, {Dwek}, {Dessart}, {Filippenko}, {Brink},
  {Foley}, {Jencson}, {Pierel}, {Zs{\'\i}ros}, {Rest}, {Zheng}, {Andrews},
  {Clayton}, {De}, {Engesser}, {Gezari}, {Gomez}, {Gonzaga}, {Johansson},
  {Kasliwal}, {Lau}, {De Looze}, {Marston}, {Milisavljevic}, {O'Steen},
  {Siebert}, {Skrutskie}, {Smith}, {Strolger}, {Van Dyk}, {Wang}, {Williams},
  {Williams}, {Xiao}, \& {Yang}}]{shahbandeh23}
{Shahbandeh}, M., {Sarangi}, A., {Temim}, T., {et~al.} 2023, \mnras, 523, 6048

\bibitem[{{Silvia} {et~al.}(2010){Silvia}, {Smith}, \& {Shull}}]{silvia10}
{Silvia}, D.~W., {Smith}, B.~D., \& {Shull}, J.~M. 2010, \apj, 715, 1575

\bibitem[{{Slavin} {et~al.}(2020){Slavin}, {Dwek}, {Mac Low}, \&
  {Hill}}]{slavin20}
{Slavin}, J.~D., {Dwek}, E., {Mac Low}, M.-M., \& {Hill}, A.~S. 2020, \apj,
  902, 135

\bibitem[{{Sluder} {et~al.}(2018){Sluder}, {Milosavljevi{\'c}}, \&
  {Montgomery}}]{sluder18}
{Sluder}, A., {Milosavljevi{\'c}}, M., \& {Montgomery}, M.~H. 2018, \mnras,
  480, 5580

\bibitem[{{Smith} {et~al.}(2009){Smith}, {Rudnick}, {Delaney}, {Rho}, {Gomez},
  {Kozasa}, {Reach}, \& {Isensee}}]{smith09}
{Smith}, J.~D.~T., {Rudnick}, L., {Delaney}, T., {et~al.} 2009, \apj, 693, 713

\bibitem[{{Smith} {et~al.}(2007){Smith}, {Draine}, {Dale}, {Moustakas},
  {Kennicutt}, {Helou}, {Armus}, {Roussel}, {Sheth}, {Bendo}, {Buckalew},
  {Calzetti}, {Engelbracht}, {Gordon}, {Hollenbach}, {Li}, {Malhotra},
  {Murphy}, \& {Walter}}]{smithJD07midIR}
{Smith}, J.~D.~T., {Draine}, B.~T., {Dale}, D.~A., {et~al.} 2007, \apj, 656,
  770

\bibitem[{{Spitzer}(1978)}]{spitzer78}
{Spitzer}, L. 1978, {Physical processes in the interstellar medium},
  doi:10.1002/9783527617722

\bibitem[{{Spyromilio} {et~al.}(1988){Spyromilio}, {Meikle}, {Learner}, \&
  {Allen}}]{spyromilio88}
{Spyromilio}, J., {Meikle}, W.~P.~S., {Learner}, R.~C.~M., \& {Allen}, D.~A.
  1988, \nat, 334, 327

\bibitem[{{Sturm} {et~al.}(2023){Sturm}, {McClure}, {Beck}, {Harsono},
  {Bergner}, {Dartois}, {Boogert}, {Chiar}, {Cordiner}, {Drozdovskaya},
  {Ioppolo}, {Law}, {Linnartz}, {Lis}, {Melnick}, {McGuire}, {Noble},
  {{\"O}berg}, {Palumbo}, {Pendleton}, {Perotti}, {Pontoppidan}, {Qasim},
  {Rocha}, {Terada}, {Urso}, \& {van Dishoeck}}]{sturm23}
{Sturm}, J.~A., {McClure}, M.~K., {Beck}, T.~L., {et~al.} 2023, \aap, 679, A138

\bibitem[{{Tinyanont} {et~al.}(2019){Tinyanont}, {Kasliwal}, {Krafton}, {Lau},
  {Rho}, {Leonard}, {De}, {Jencson}, {Mawet}, {Millar-Blanchaer}, {Nilsson},
  {Yan}, {Gehrz}, {Helou}, {Van Dyk}, {Serabyn}, {Fox}, \&
  {Clayton}}]{tinyanont19}
{Tinyanont}, S., {Kasliwal}, M.~M., {Krafton}, K., {et~al.} 2019, \apj, 873,
  127

\bibitem[{{Todini} \& {Ferrara}(2001)}]{todini01}
{Todini}, P., \& {Ferrara}, A. 2001, \mnras, 325, 726

\bibitem[{{van der Tak} {et~al.}(2007){van der Tak}, {Black}, {Sch{\"o}ier},
  {Jansen}, \& {van Dishoeck}}]{vanderTak07}
{van der Tak}, F.~F.~S., {Black}, J.~H., {Sch{\"o}ier}, F.~L., {Jansen}, D.~J.,
  \& {van Dishoeck}, E.~F. 2007, \aap, 468, 627

\bibitem[{{Wallstr{\"o}m} {et~al.}(2013){Wallstr{\"o}m}, {Biscaro}, {Salgado},
  {Black}, {Cherchneff}, {Muller}, {Bern{\'e}}, {Rho}, \&
  {Tielens}}]{wallstrom13}
{Wallstr{\"o}m}, S.~H.~J., {Biscaro}, C., {Salgado}, F., {et~al.} 2013, \aap,
  558, L2 (W13)

\bibitem[{{Young} {et~al.}(2006){Young}, {Fryer}, {Hungerford}, {Arnett},
  {Rockefeller}, {Timmes}, {Voit}, {Meakin}, \& {Eriksen}}]{young06}
{Young}, P.~A., {Fryer}, C.~L., {Hungerford}, A., {et~al.} 2006, \apj, 640, 891

\end{thebibliography}

\end{document}